\pdfoutput=1
\documentclass[a4paper,USenglish,cleveref]{lipics-v2021}

\usepackage[normalem]{ulem}

\usepackage{amsmath,amsfonts}
\usepackage{algorithmic}
\usepackage{graphicx}
\usepackage{textcomp}
\usepackage{xcolor}
\usepackage{amsthm}
\usepackage{url}
\usepackage[n,
            advantage,
            operators,
            sets,
            adversary,
            landau,
            probability,
            notions,
            logic,
            ff,
            mm,
            primitives,
            events,
            complexity,
            asymptotics,
            keys]{cryptocode}
\usepackage{tikz}
\usetikzlibrary{arrows.meta}
\usetikzlibrary{patterns}
\usetikzlibrary{decorations.pathreplacing}

\usepackage{xspace}
\newcommand{\wlist}{AllowList\xspace}
\newcommand{\blist}{DenyList\xspace}
\newcommand{\wlists}{AllowLists\xspace}
\newcommand{\blists}{DenyLists\xspace}

\newboolean{showcomments}
\setboolean{showcomments}{true}
\ifthenelse{\boolean{showcomments}}
{ \newcommand{\annote}[3]{{
		\colorbox{#3}{\bfseries\sffamily\footnotesize\textcolor{white}{#2}}
		\color{#3}
		$\blacktriangleright${\it #1}$\blacktriangleleft$}
}
}

\newcommand{\df}[1]{\annote{#1}{DF}{red}}

\DeclareRobustCommand{\change}[4]{{\color{#3}#1}
\textcolor{gray}{\sout{#4}}\ftanote{\color{#3} \sc}{#2}}

\DeclareRobustCommand{\chadd}[3]{{\color{#3}#1\ftanote{\color{#3} \sc}{#2}}}
\newcommand{\MGa}[1]{\chadd{#1}{MG}{orange}}
\newcommand{\DF}[2]{\change{#1}{DF}{red}{#2}}

\renewcommand{\chadd}[3]{#1}
\renewcommand{\change}[4]{#1}
\renewcommand{\annote}[3]{}

\usepackage{newfloat}
\DeclareFloatingEnvironment[fileext=frm,placement={!ht},name=Algorithm]{algorithm}

\title{The Synchronization Power (Consensus Number) of  Access-Control Objects: The Case of AllowList and DenyList}
\titlerunning{The Synchronization Power of Access Control Objects}
\author{Davide Frey}{Inria, IRISA, CNRS, Université de Rennes}{davide.frey@inria.fr}{}{}
\author{Mathieu Gestin}{Inria, IRISA, CNRS, Université de Rennes}{mathieu.gestin@inria.fr}{}{}
\author{Michel Raynal}{IRISA, Inria, CNRS, Université de Rennes}{michel.raynal@irisa.fr}{}{}

\authorrunning{Davide Frey, Mathieu Gestin, Michel Raynal}

\Copyright{Davide Frey, Mathieu Gestin, and Michel Raynal}



\EventEditors{Rotem Oshman}
\EventNoEds{1}
\EventLongTitle{37th International Symposium on Distributed Computing (DISC 2023)}
\EventShortTitle{DISC 2023}
\EventAcronym{DISC}
\EventYear{2023}
\EventDate{October 10-12, 2023}
\EventLocation{L'Aquila, Italy}
\EventLogo{}
\SeriesVolume{281}
\ArticleNo{39}

\funding{This work was partially funded by the PriCLeSS project and by the SOTERIA H2020 project. PriCLeSS
was granted by the Labex CominLabs excellence laboratory of the French ANR (ANR-10-LABX-07-01).
SOTERIA received funding from the European Union’s Horizon 2020 research and innovation programme
under grant agreement No101018342. This content reflects only the author’s view. The European
Agency is not responsible for any use that may be made of the information it contains.}

\acknowledgements{We wish to thank the anonymous reviewers for their insightful comments and remarks that
led to significant improvements to our paper.}

\begin{document}
\nolinenumbers
\maketitle
% ----------------------------------------------------------------

%\setcopyright{none}
%\copyrightyear{2022}
%\acmYear{2022}
%\acmDOI{}
%\acmConference[]{}{}{}
%\acmBooktitle{}
%\acmPrice{}
%\acmISBN{}
\begin{abstract}
This article studies the synchronization power of \wlist and \blist objects
under the lens provided by Herlihy's consensus hierarchy. It specifies \wlist and
\blist as distributed objects and shows that, while they can both be
seen as specializations of a more general object type, they
inherently have different synchronization power.  While the
\wlist object does not require synchronization between participating
processes, a \blist object requires processes to reach
consensus on a specific set of processes. These results
are then applied to a more global analysis of anonymity-preserving systems that use \wlist and
\blist objects. The specification .First, a blind-signature-based e-voting is presented. Second, DenyList and AllowList objects are used to determine
the consensus number of a specific decentralized key management system. Third, an anonymous money
transfer protocol using the association of AllowList and DenyList objects is presented. Finally, this study is used
to study the properties of these application, and to highlight efficiency gains that they can achieve in message passing environment.
\end{abstract}

\section{Introduction}
%virer blockchain, parler que des implems et de ce qui est le problèmes.
The advent of blockchain technologies increased the interest of the
public and industry in distributed applications, giving birth to
projects that have applied blockchains in a plethora of use
cases. These include e-vote systems~\cite{Bronco}, naming
services~\cite{ENS,Namecoin}, Identity Management
Systems~\cite{Sovrin,uport}, supply-chain
management~\cite{block-supply-chain}, or Vehicular Ad hoc Network
\cite{block-vanet}. However, this use of the blockchain as a
swiss-army knife that can solve numerous distributed problems highlights a
lack of understanding of the actual requirements of those problems.
Because of these poor specifications, implementations of these applications are
often sub-optimal.

This paper thoroughly studies
a class of problems widely used in distributed applications and provides a
guideline to implement them with reasonable but sufficient tools.

Differently from the previous approaches, it aims to understand the amount of synchronization required between processes of a system
to implement \textit{specific} distributed objects. To achieve this goal it studies such objects under the lens of Herlihy's consensus number \cite{herlihy}. This parameter is inherently associated to
shared memory distributed objects, and has no direct correspondence in the message passing environment. However, in some specific cases, this information is enough to provide a better understanding of the objects analyzed, and thus, to gain efficiency in the message passing implementations. For example, recent papers~\cite{Guerraoui,Auvolat} have shown that cryptocurrencies
can be implemented without consensus and therefore without a
blockchain. In particular, Guerraoui et al.~\cite{Guerraoui} show that
$k$-asset transfer has a consensus number $k$ where
$k$ is the number of processes that can withdraw currency from the
same account~\cite{cons-number}. Similarly, Alpos et al.~\cite{Zanolini} have studied the
synchronization properties of ERC20 token smart contracts and shown
that their consensus number varies over time as a result of
changes in the set of processes that are approved to send tokens from
the same account. These two results consider two forms of asset
transfer: the classical one and the one implemented
by the ERC20 token, which allows processes to
dynamically authorize other processes. The consensus number of those objects
depends on specific and well identified processes. From this study, it is possible to
conclude that the consensus algorithms only need to be performed between those processes.
Therefore, in these specific cases, the knowledge of the consensus number of an object can
be directly used to implement more efficient message passing applications. Furthermore, even if
this study uses a shared memory model, with crash prone processes, its results can
be used to implement more efficient Byzantine resilient algorithm, in a message passing
environment.
This paper proposes to extend this knowledge to a broader class of applications.

Indeed, the transfer of assets, be them cryptocurrencies or non-fungible
tokens, does not constitute the only application in the Blockchain
ecosystem. In particular, as previously indicated, a number of applications like
e-voting~\cite{Bronco}, naming~\cite{ENS,Namecoin}, or Identity
Management~\cite{Sovrin,uport} use Blockchain as a tool to implement
some form of access control. This is often achieved by implementing
two general-purpose objects: \wlists and \blists. An \wlist provides
an opt-in mechanism. A set of managers can maintain a list of
authorized parties, namely the \wlist. To access a resource, a party (user)
must prove the presence of an element associated with its identity in the \wlist. A
\blist provides instead an opt-out mechanism. In this case, the
managers maintain a list of revoked elements, the \blist. To access a
resource, a party (user) must prove that no corresponding element
has been added to the \blist. In other words, \wlist and \blist
support, respectively, set-membership and set-non-membership proofs on
a list of elements.

The proofs carried out by \wlist and \blist objects often need to
offer privacy guarantees. For example, the Sovrin privacy preserving Decentralized
Identity-Management System (DIMS)~\cite{Sovrin} associates an
\wlist\footnote{In reality this is a variant that mixes \wlist
  and \blist which we discuss in \cref{sec:variations}.} with each
verifiable credential that contains the identifiers of the devices
that can use this verifiable credential. When a device uses a
credential with a verifier, it needs to prove that the identifier
associated with it belongs to the \wlist. This proof must be done in
zero knowledge, otherwise the verifier would learn the identity of the
device, which in turn could serve as a pseudo-identifier for the
user. For this reason, \wlist and \blist objects support
respectively a zero-knowledge proof of set membership or a
zero-knowledge proof of set non-membership.

Albeit similar, the \wlist and \blist objects differ significantly in
the way they handle the proving mechanism.
In the case of an \wlist, no security risk appears if access to a
resource is prohibited to a process, even if a manager did grant this right.
 As a result, a transient period in which a user is first allowed, then denied, and then allowed again to access a resource poses no problem. On the
contrary, with a \blist, being allowed access to a resource after being
denied poses serious security problems. Hence, the \blist object is defined with
an additional anti-flickering property prohibiting those transient periods.
This property is the main difference between an \wlist and a \blist object and is the reason for their distinct consensus numbers.

Existing systems~\cite{Bronco,ENS,Namecoin,Sovrin,uport} that employ
\wlist and \blist objects implement them on top of a heavy blockchain
infrastructure, thereby requiring network-level consensus to modify
their content.
As already said, this paper studies this difference under the lens of the consensus
number~\cite{cons-number}. It shows that (i) the consensus number of an
\wlist object is $1$, which means that an \wlist can be implemented
without consensus; and that (ii) the consensus number of a \blist is instead
equal to the number of processes that can conduct prove operations on the
\blist, and that only these processes need to synchronize. Both data structures can
therefore be implemented without relying on the network-level
consensus provided by a blockchain, which opens the door to more efficient
implementations of applications based on these data structures.

To summarize, this paper presents the following three contributions.
\begin{enumerate}
\item It formally defines and studies \wlist and \blist as distributed objects (\Cref{sec:definitions}).
\item It analyses the consensus number of these objects: it shows that the \wlist does not require synchronization between processes (\Cref{sec:wlist}), while the \blist requires the synchronization of all the verifiers of its set-non-membership proofs (\Cref{sec:blist}).
\item It uses these theoretical results to give intuitions on
  their optimal implementations.
  Namely the implementation of a DIMS, as well as of an e-vote system and an
  anonymous asset-transfer protocol (\Cref{sec:aat} and \ref{sec:evote}).
\end{enumerate}
% \begin{center}
%   \includegraphics[width=8cm]{outline.pdf}
% \end{center}
% \vspace{1ex}
To the best of our knowledge, this paper is the first to study the \wlist and \blist from a distributed algorithms point of view. So we believe our results can provide a powerful tool to identify the consensus number of recent distributed objects that make use of them and to provide more
efficient implementations of such objects.

\section{Preliminaries}\label{sec:preliminaries}
\subsection{Computation Model}\label{sec:computmodel}
\paragraph*{Model}
Let $\Pi$ be a set of $N$ asynchronous sequential crash-prone processes $p_1, \cdots, p_N$. Sequential means that each process invokes one operation of its own algorithm at a time. We assume the local processing time to be instantaneous, but the system is asynchronous. This means that non-local operations can take a finite but arbitrarily long time and that the relative speeds between the clocks of the different processes are unknown. Finally, processes are crash-prone: any number of processes can prematurely and definitely halt their executions. A process that crashes is called \textit{faulty}. Otherwise, it is called \textit{correct}. The system is eponymous: a unique positive integer identifies each process, and this identifier is known to all other processes.

\paragraph*{Communication}
Processes communicate via shared objects of type $T$. Each operation on a shared object is associated with two \textit{events}: an \textit{invocation} and a \textit{response}. An object type $T$ is defined by a tuple $(Q, Q_0, O, R, \Delta)$, where $Q$ is a set of states, $Q_0 \subseteq Q$ is the set of initial states, $O$ is the set of operations a process can use to access this object, $R$ is the set of responses to these operations, and $\Delta \subseteq \Pi \times Q  \times O \times R \times Q$ is the transition function defining how a process can access and modify an object.\df{This is the definition of a relation, not a function isn't it?}

\paragraph*{Histories and Linearizability}
%To prove the correctness of our algorithms, we will use Herlihy and Wing \cite{herlihy} definition of linearizability for concurrent objects.
A \textit{history}~\cite{herlihy} is a sequence of invocations and
responses in the execution of an algorithm.
% The \textit{execution} $E$ of a multi-process shared memory
% algorithm can be represented by a sequence of invocations and
% responses which follows the prescribed algorithm. We call this
% sequence a \textit{history}.
An invocation with no matching response in a history, $H$, is called a \textit{pending} invocation. A \textit{sequential history} is one where the first event is an invocation, and each
invocation---except possibly the last one---is immediately
followed by the associated response. A sub-history is a sub-sequence of events in a history. A process sub-history $H|
p_i$ of a history $H$ is a sub-sequence of all the events in $H$
whose associated process is $p_i$. Given an object $x$, we can similarly define the object sub-history $H|x$. Two histories $H$ and $H'$ are equivalent if $H|p_i = H'|p_i$, $\forall i \in \{1, \cdots, N\}$. %Because our processes are
%sequential, all histories we will study are well-formed, i.e., for
%a history $H$ and a given process $p_i$, $H|p_i$ is sequential.

In this paper, we define the specification of a shared object, $x$, as
the set of all the allowed sub-histories, $H|x$. We talk about a
sequential specification if all the histories in this set are
sequential. A \textit{legal history} is a history $H$ in which, for all
objects $x_i$ of this history, $H|x_i$ belongs to the \df{sequential?}
specification of $x_i$.  The completion $\bar{H}$ of a history $H$ is
obtained by extending all the pending invocations in $H$ with the
associated matching responses.
%Given a history $H$, we can define a completion $\bar{H}$ of this history by extending all pending invocations in $H$ with the associated matching responses.
A history $H$ induces an irreflexive partial order $<_H$ on
operations, i.e. $op_0 <_H op_1$ if the response to the operation $op_0$
precedes the invocation of operation $op_1$. A history is sequential if
$<_H$ is a total order. The algorithm executed by a correct process is
\textit{wait-free} if it always terminates after a finite number of
steps.  A history $H$ is linearizable if a completion $\bar{H}$ of $H$
is equivalent to some legal sequential history $S$ and
$<_H \subseteq <_S$.

\paragraph*{Consensus number}
The consensus number of an object of type $T$ (noted cons($T$)) is the largest $n$ such that it is possible to wait-free implement a consensus object from atomic read/write registers and objects of type $T$ in a system of $n$ processes. If an object of type $T$ makes it possible to wait-free implement a consensus object in a system of any number of processes, we say the consensus number of this object is $\infty$. Herlihy~\cite{cons-number} proved the following well-known theorem.
\begin{theorem}\label{theorem: consensus}
	Let $X$ and $Y$ be two atomic objects type such that cons($X) = m$ and cons($Y)=n$, and $m<n$. There is no wait-free implementation of an object of type $Y$ from objects of type $X$ and read/write registers in a system of more than $m$ processes.
\end{theorem}

We will determine the consensus number of the \blist and the \wlist objects using Atomic Snapshot objects and consensus objects in a set of $k$ processes. A Single Writer Multi Reader (SWMR)~\cite{Afek} Atomic Snapshot object is an array of fixed size, which supports two operations: Snapshot and Update. The Snapshot() operation allows a process $p_i$ to read the whole array in one atomic operation. The Update($v$, $i$) operation allows a process $p_i$ to write the value $v$ in the $i$-th position of the array. Afek et al. showed that a SWMR Snapshot object can be wait-free implemented from read/write registers \cite{Afek}, i.e., this object type has consensus number $1$. This paper assumes that all Atomic Snapshot objects used are SWMR.
A consensus object provides processes with a single one-shot operation \emph{propose()}. When a process $p_i$ invokes \emph{propose(v)} it proposes $v$. This invocation returns a \emph{decided} value such that the following three properties are satisfied.
 \begin{itemize}
 \item \textit{Validity}: If a correct process decides value $v$, then $v$ was proposed by some process;
 \item \textit{Agreement}: No two correct processes decide differently; and
 \item \textit{Termination}: Every correct process eventually decides.
 \end{itemize}
 A $k$-consensus object is a consensus object accessed by at most $k$ processes. % takes a proposed value as input and returns a decided value. There can be up to $k$ accesses to a given $k$-consensus object, any extra access will return $\emptyset$. The properties of this object are:

\subsection{Number theory preliminaries}\label{sec:numbertheory}
\paragraph*{Cryptographic Commitments}
%In distributed systems, when there is no Trusted Third Party
A \textit{cryptographic commitment} is a cryptographic scheme that allows a Prover to commit to a value $v$ while hiding it. The commitment scheme is a two phases protocol. First, the prover computes a binding value known as commitment, $C$, using a function \textit{Commit}. \textit{Commit} takes as inputs the value $v$ and a random number $r$. The prover sends this hiding and binding value $C$ to a verifier. In the second phase, the prover reveals the committed value $v$ and the randomness $r$ to the verifier. The verifier can then verify that the commitment $C$ previously received refers to the transmitted values $v$ and $r$. This commitment protocol is the heart of Zero Knowledge Proof (ZKP) protocols.

\paragraph*{Zero Knowledge Proof of set operations}
A Zero Knowledge Proof (ZKP) system is a cryptographic protocol that allows a prover to prove some Boolean statement about a value $x$ to a verifier without leaking any information about $x$. A ZKP system is initialized for a specific language $\mathcal{L}$ of the complexity class $\mathcal{NP}$. The proving mechanism takes as input $\mathcal{L}$  and outputs a proof $\pi$. Knowing $\mathcal{L}$ and $\pi$, any verifier can verify that the prover knows a value $x \in \mathcal{L}$\footnote{The notation $x \in \mathcal{L}$ denotes the fact that $x$ is a solution to the instance of the problem expressed by the language $\mathcal{L}$}. However, the verifier cannot learn the value $x$ used to produce the proof. In the following, it is assumed there exists efficient non interactive ZKP systems of set-membership and set-non-membership (e.g., constructions from \cite{ZK} can be used).

%In our case, we will only use two types of languages. The language of all elements that belongs to a predefined set $\mathcal{S}$, and the language of all elements that do not belongs to another predefined set $\mathcal{S}'$. The first language represents proofs of set-membership and will be used to implement distributed \wlist objects. The second language represents proof of set-non-membership and will be used to implement distributed \blist objects. %Efficient ZKP systems have been proposed for proof of set-membership and proof of set-non-membership using accumulators~\cite{BCFK19}, details about cryptographic accumulators are given in \Cref{sec:sync}.
\section{The \wlist and \blist objects: Definition}\label{sec:definitions}
Distributed \wlist and \blist object types are the type of objects that allow a set of managers to control access to a resource. The term ''resource'' is used here to describe the goal a user wants to achieve and which is protected by an access control policy. A user is granted access to the resource if it succeeds in proving that it is authorized to access it. First, we describe the \wlist object type. Then we consider the \blist object type.

The \wlist object type is one of the two most common access control mechanisms. To access a resource, a process \MGa{$p \in \Pi_V$} needs to prove it knows some element $v$ previously authorized by a process $p_M \in \Pi_M$, where $\Pi_M \subseteq \Pi$ is the set of managers, and $\Pi_V \subseteq \Pi$ is the set of processes authorized to conduct proofs. We call verifiers the processes in $\Pi_V$. The sets $\Pi_V$ and $\Pi_M$ are predefined and static. They are parameters of the object. Depending on the usage of the object, these subset can either be small, or they can contain all the processes in $\Pi$.

A process $p\in \Pi_V$ proves that $v$ was previously authorized by invoking a PROVE($v$) operation. This operation is said to be valid if some manager in $\Pi_M$ previously invoked an APPEND($v$) operation. Intuitively, we can see the invocation of the APPEND($v$) operation as the action of authorizing some process to access the resource.
On the other hand, the PROVE($v$) operation, performed by a prover process, $p\in \Pi_V$, proves to the other processes in $\Pi_V$ that they are authorized. However, this proof is not enough in itself. The verifiers of a proof must be able to verify that a valid PROVE operation has been invoked. To this end, the \wlist object type is also equipped with a READ() operation. This operation can be invoked by any process in $\Pi$ and returns all the valid PROVE operations invoked, along with the identity of the processes that invoked them. The list returned by the READ operation can be any arbitrary permutation of the list of PROVE operations. All processes in $\Pi$ can invoke the READ operation.\footnote{Usually, AllowList objects are implemented in a message-passing setting. In these cases, the READ operation is implicit. Each process knows a local state of the distributed object, and can inspect it any time. In the shared-memory setting, we need to make this READ operation explicit.}

An optional anonymity property can be added to the \wlist object to enable privacy-preserving implementations. This property ensures that other processes cannot learn the value $v$ proven by a PROVE($v$) operation.

The \wlist object type is formally defined as a sequential object, where each invocation is immediately followed by a response. Hence, the sequence of operations defines a total order, and each operation can be identified by its place in the sequence.

\begin{definition}
The \emph{\wlist} object type supports three operations: \normalfont{APPEND, PROVE, and READ}. These operations appear as if executed in a sequence $\mathsf{Seq}$ such that:
\begin{itemize}
	\item \textit{Termination.} A \normalfont{PROVE}, an \normalfont{APPEND}, or a \normalfont{READ} operation invoked by a correct process always returns.
	\item APPEND \textit{Validity.} The invocation of APPEND($x$) by a process $p$ is valid \textbf{if}:
		\begin{itemize}
			\item $p \in \Pi_M \subseteq \Pi$;\textbf{ and}
			 \item $x \in \mathcal{S}$, where $\mathcal{S}$ is a predefined set.
		\end{itemize}
		Otherwise, the operation is invalid.
	\item PROVE \textit{Validity.} \textbf{If} the invocation of $op=$\normalfont{PROVE}($x$) by a process $p$ is valid, \textbf{then}:
		\begin{itemize}
			\item $p \in \Pi_V \subseteq \Pi$; \textbf{and}
			%\item $x \in \mathcal{S}$\footnote{This condition is actually implied by the APPEND validity and the next condition. It is made explicitly here for clarity.}; and
			\item A valid \normalfont{APPEND}($x$) operation appears before $op$ in $\mathsf{Seq}$.
		\end{itemize}
		Otherwise, the invocation is invalid.
		\item \textit{Progress.} \textbf{If }a valid \normalfont{APPEND}($x$) operation is invoked, \textbf{then } there exists a point in $\mathsf{Seq}$ such that any \normalfont{PROVE}($x$) operation invoked after this point by any process $p \in \Pi_V$ will be valid.
	\item READ \textit{Validity.} The invocation of $op=$\normalfont{READ}() by a process $p \in \Pi_V$ returns the list of valid invocations of \normalfont{PROVE} that appears before $op$ in $\mathsf{Seq}$ along with the names of the processes that invoked each operation.
	\item \textit{Optional - Anonymity.} Let us assume the process $p$ invokes a PROVE($v$) operation. If the process $p'$ invokes a READ() operation, then $p'$ cannot learn the value $v$ unless $p$ leaks additional information.\footnote{The Anonymity property only protects the value $v$. The system considered is eponymous. Hence, the identity of the processes is already known. However, the anonymity of $v$ makes it possible to hide other information. For example, the identity of a client that issues a request to a process of the system. These example are discussed in \Cref{sec:discussion}. Thereby, the anonymity property does not contravene the READ validity property, which only discloses the process identity.}
\end{itemize}
\end{definition}

The \wlist object is defined in an append-only manner. This definition makes it possible to use it to build all use cases explored in this paper. However, some use cases could need an \blist with an additional REMOVE operation.
This variation is studied in \Cref{sec:variations}.
%\Cref{fig:wlist} gives an example execution of an \wlist object, where $\Pi = \Pi_V = \Pi_M = \{p_1, p_2\}$. The PROVE and the APPEND validity are highlighted, along with the progress property. A return value of \emph{True} means the operation is valid, and one of \emph{False} means it is not. In the figure, a ''flickering'' zone appears. This behavior is made possible by the PROVE validity property. In this zone, two processes can see different responses after the invocation of the same operation. It is the main difference between the \wlist and the \blist object types. %In the ''progress'' zone and the ''flickering'' zone, two READ operations can return completely different sets.
%\begin{figure}
%\center
%\includegraphics[scale=0.3]{img/whitelist_prez.png}
%\Description[Example execution of an \wlist object.]{Example execution of an \wlist object.}
%\caption{Example execution of an \wlist object.}
%\label{fig:wlist}
%\end{figure}

The \blist object type can be informally presented as an access policy where, contrary to the \wlist object type, all users are authorized to access the resource in the first place. The managers are here to revoke this authorization. A manager revokes a user by invoking the APPEND($v$) operation. A user uses the PROVE($v$) operation to prove that it was not revoked. A PROVE($v$) invocation is invalid only if a manager previously revoked the value $v$.

All the processes in $\Pi$ can verify the validity of a PROVE operation by invoking a READ() operation. This operation is similar to the \wlist's READ operation. It returns the list of valid PROVE invocations along with the name of the processes that invoked it.

There is one significant difference between the \blist and the \wlist object types. With an \wlist, if a user cannot access a resource immediately after its authorization, no malicious behavior can harm the system---the system's state is equivalent to its previous state. However, with a \blist, a revocation not taken into account can let a malicious user access the resource and harm the system. In other words, access to the resource in the \blist case must take into account the ''most up to date'' \DF{available revocation list}{list of revocation available}.

To this end, the \blist object type is defined with an additional property. The anti-flickering property ensures that if an APPEND operation is taken into account by one PROVE operation, it will be taken into account by every subsequent PROVE operation. Along with the progress property, the anti-flickering property ensures that the revocation mechanism is as immediate as possible. The \blist object is formally defined as a sequential object, where each invocation is immediately followed by a response. Hence, the sequence of operations define a total order, and each operation can be identified by its place in the sequence.

\begin{definition}
The \emph{\blist} object type supports three operations: \normalfont{APPEND, PROVE, and READ}. These operations appear as if executed in a sequence $\mathsf{Seq}$ such that:
\begin{itemize}
	\item \textit{Termination.} A \normalfont{PROVE}, an \normalfont{APPEND}, or a \normalfont{READ} operation invoked by a correct process always returns.
	\item APPEND \textit{Validity.} The invocation of \normalfont{APPEND}($x$) by a process $p$ is valid \textbf{if}:
		\begin{itemize}
			\item $p \in \Pi_M \subseteq \Pi$; \textbf{and}
			 \item $x \in \mathcal{S}$, where $\mathcal{S}$ is a predefined set.
		\end{itemize}
	Otherwise, the operation is invalid.
	\item PROVE \textit{Validity.} \textbf{If} the invocation of a $op=$\normalfont{PROVE}($x$) by a correct process $p$ is not valid, \textbf{then}:
		\begin{itemize}
			\item $p \notin \Pi_V \subseteq \Pi$; \textbf{or}
			%\item $x \notin \mathcal{S}$; or
			\item A valid \normalfont{APPEND}$(x)$ appears before $op_P$ in $\mathsf{Seq}$.
		\end{itemize}
		Otherwise, the operation is valid.
	\item PROVE \textit{Anti-Flickering.} \textbf{If} the invocation of a operation $op=$\normalfont{PROVE}($x$) by a correct process $p \in \Pi_V$ is invalid, \textbf{then }any \normalfont{PROVE}($x$) operation that appears after $op$ in $\mathsf{Seq}$ is invalid.\footnote{The only difference between the \wlist and the \blist object types is this anti-flickering property. As it is shown in \Cref{sec:wlist} and in \Cref{sec:blist}, the \wlist object has consensus number 1, and the \blist object has consensus number $k=|\Pi_V|$. Hence, this difference in term of consensus number is due solely to the anti-flickering property. It is an open question whether a variation of this property could transform any consensus number $1$ object into a consensus number $k$ object.}
		\item READ \textit{Validity.} The invocation of $op=$\normalfont{READ}() by a process $p \in \Pi_V$ returns the list of valid invocations of \normalfont{PROVE} that appears before $op$ in $\mathsf{Seq}$ along with the names of the processes that invoked each operation.
	\item \textit{Optional - Anonymity.} Let us assume the process $p$ invokes a PROVE($v$) operation. If the process $p'$ invokes a READ() operation, then $p'$ cannot learn the value $v$ unless $p$ leaks additional information.
\end{itemize}
\end{definition}

%\Cref{fig:blist} gives an example execution of a \blist object, where $\Pi = \Pi_V = \Pi_M = \{p_1, p_2\}$. The PROVE and the APPEND validity are highlighted, along with the anti-flickering and the progress properties. In the ''progress + anti-flickering'' zone, any READ will return the same value independently of the process that invokes the operation.
%\begin{figure}
%\center
%\includegraphics[scale=0.3]{img/blacklist_prez.png}
%\Description[Example execution of a \blist object.]{Example execution of a \blist object.}
%\caption{Example Execution of a \blist object.}
%\label{fig:blist}
%\end{figure}
\section{PROOF-LIST object specification}
\label{sec:prooflist}
\begin{table}
\resizebox{14cm}{!}{
\begin{tabular}{l|l|l|l|l|l}
	Process & Operation & Initial state & Res- & Final state & Conditions\\
	 	&	&	& ponse & & \\\hline
	$p_i \in \Pi_M$ & APPEND($y$) & $(\textit{listed-values} = \{x\in \mathcal{S}\}$, & True & $(\textit{listed-values} \cup \{y\},$ & $y \in \mathcal{S}$ \\
	& & $\textit{proofs} = (\{(p_j \in \Pi, \widehat{\mathcal{S}} \subseteq \mathcal{S}, \mathsf{P}\in \mathcal{P}_{\mathcal{L}_{\widehat{\mathcal{S}}}})\}))$& &\textit{proofs})\\
	$p_i$ & APPEND($y$) & $(\textit{listed-values} = \{x\in \mathcal{S}\},$& False & $(\textit{listed-values}, \textit{proofs})$ & $p_i \notin \Pi_M \lor y \notin \mathcal{S}$ \\
	& & $\textit{proofs} = (\{(p_j \in \Pi, \widehat{\mathcal{S}} \subseteq \mathcal{S}, \mathsf{P}\in \mathcal{P}_{\mathcal{L}_{\widehat{\mathcal{S}}}})\}))$& &\\
	$p_i \in \Pi_V$ & PROVE($y$) & $(\textit{listed-values} = \{x\in \mathcal{S}\},$ & $(\mathcal{A}, \mathsf{P})$ & $(\textit{listed-values},$ & $\forall y \in \mathcal{L}_{\mathcal{A}} \wedge \mathcal{A} \subseteq \textit{listed-values}$ \\
	& & $\textit{proofs} = (\{(p_j \in \Pi, \widehat{\mathcal{S}} \subseteq \mathcal{S}, \mathsf{P}\in \mathcal{P}_{\mathcal{L}_{\widehat{\mathcal{S}}}})\}))$& &$\textit{proofs} \cup \{(p_i, \mathcal{A}, \mathsf{P})\})$ &$\wedge \forall \mathsf{P} \in \mathcal{P}_{\mathcal{L}_{\mathcal{A}}} \wedge \mathsf{C}(y, \widehat{\mathcal{S}})=1$\\
	$p_i$ & PROVE($y$) & $(\textit{listed-values} = \{x\in \mathcal{S}\}$, & False & $(\textit{listed-values}, \textit{proofs})$ & $ \forall y \notin \mathcal{L}_{\mathcal{A}} \lor \mathcal{A} \not \subseteq \textit{listed-values}$ \\
	& & $\textit{proofs} = (\{(p_j \in \Pi, \widehat{\mathcal{S}} \subseteq \mathcal{S}, \mathsf{P}\in \mathcal{P}_{\mathcal{L}_{\widehat{\mathcal{S}}}})\}))$& & &$\lor\ \forall \mathsf{P} \notin \mathcal{P}_{\mathcal{L}_{\mathcal{A}}} \lor \forall p_i \notin \Pi_V$\\
	& & & & & $\lor\ \mathsf{C}(y, \widehat{\mathcal{S}})=0$\\
	$p_i \in \Pi$ & READ() & $(\textit{listed-values} = \{x\in \mathcal{S}\},$ & $\textit{proofs}$ & $(\textit{listed-values}, \textit{proofs})$ &  \\
	& & $\textit{proofs} = (\{(p_j \in \Pi, \widehat{\mathcal{S}} \subseteq \mathcal{S}, \mathsf{P}\in \mathcal{P}_{\mathcal{L}_{\widehat{\mathcal{S}}}})\}))$& &\\
\end{tabular}
}
\caption{Transition function $\Delta$ for the PROOF-LIST object.}
\label{fig:transition function}
\end{table}
\Cref{sec:wlist} and \Cref{sec:blist} propose an analysis of the synchronization power of the \wlist and the \blist object types using the notion of consensus number. Both objects share many similarities. Indeed, the only difference is the type of proof performed by the user and the non-flickering properties. Therefore, this section defines the formal specification of the PROOF-LIST object type, a new generic object that can be instantiated to describe the \wlist or the \blist object type.

The PROOF-LIST object type is a distributed object type whose state is a pair of arrays (\textit{listed-values}, \textit{proofs}). The first array, \textit{listed-values}, represents the list of authorized/revoked elements. It is an array of objects in a set $\mathcal{S}$, where $\mathcal{S}$ is the universe of potential elements. The second array, \textit{proofs}, is a list of assertions about the \textit{listed-values} array. Given a set of managers $\Pi_M \subseteq \Pi$ and a set of verifiers $\Pi_V \subseteq \Pi$, the PROOF-LIST object supports three operations. First, the APPEND($v$) operation appends a value $v \in \mathcal{S}$ to the \textit{listed-values} array. Any process in the manager's set can invoke this operation. Second, the PROVE($v$) operation appends a valid proof about the element $v \in \mathcal{S}$ relative to the \textit{listed-values} array to the \textit{proofs} array. This operation can be invoked by any process $p \in \Pi_V$. Third, the READ() operation returns the \textit{proofs} array.

The sets $\Pi_V$ and $\Pi_M$ are static, predefined subsets of $\Pi$. There is no restriction on their compositions. The choice of these sets only depends on the usage of the \wlist or the \blist. Depending on the usage, they can either contain a small subset of processes in $\Pi$
% ---this is the case with SSI, where $\Pi_V$ are the set of issuing entities, and $\Pi_M$ is defined in an ad hoc way with the verifier of a credential and the owner of the credential---
or they can contain the whole set of processes of the system.%---this case occurs with anonymous money transfer objects \Cref{sec:aat}.

To express the proofs produced by a process $p$, we use an abstract language $\mathcal{L}_{\mathcal{A}}$ of the complexity class $\mathcal{NP}$, which depends on a set $\mathcal{A}$. This language will be specified for the \wlist and the \blist objects in \Cref{sec:wlist} and \Cref{sec:blist}. The idea is that $p$ produces a proof $\pi$ about a value $v \in \mathcal{S}$. A PROVE invocation by a process $p$ is valid only if the proof $\pi$ added to the \textit{proofs} array is valid. The proof $\pi$ is valid if $v \in \mathcal{L}_{\mathcal{A}}$---i.e., $v$ is a solution to the instance of the problem expressed by $\mathcal{L}_{\mathcal{A}}$, where $\mathcal{L}_{\mathcal{A}}$  is a language of the complexity class $\mathcal{NP}$ \footnote{In this article, $\mathcal{L}_{\mathcal{A}}$ can be one of the following languages: a value $v$ belongs to $\mathcal{A}$ (\wlist), or a value $v$ does not belongs to $\mathcal{A}$ (\blist).}
which depends on a  subset ${\mathcal{A}}$ of the \textit{listed-values} array ($\mathcal{A} \subseteq \mathcal{S}$). We note $\mathcal{P}_{\mathcal{L}_{\mathcal{A}}}$ the set of valid proofs relative to the language $\mathcal{L}_{\mathcal{A}}$. $\mathcal{P}_{\mathcal{L}_{\mathcal{A}}}$ can either represent Zero Knowledge Proofs or explicit proofs.
%Because the language cannot be directly linked to the \textit{listed-values} array --- the language $\mathcal{L}$ is define before the definition of a specific $\textit{listed-values}$ array---, it is the verifiers task to verify that the set ${\widehat{\mathcal{A}}}$ is a subset of the \textit{listed-values} array.

If a proof $\pi$ is valid, then the PROVE operation returns $({{\mathcal{A}}}, \mathsf{Acc}.Prove(v, {{\mathcal{A}}}))$, where $\mathsf{Acc}.Prove(v, {{\mathcal{A}}})$ is the proof generated by the operation, and where ${{\mathcal{A}}}$ is a subset of values in \textit{listed-values} on which the proof was applied. Otherwise, the PROVE operation returns "False". Furthermore, the \textit{proofs} array also stores the name of the processes that invoked PROVE operations.

Formally, the PROOF-LIST object type is defined by the tuple ($Q$, $Q_0$, $O$, $R$, $\Delta$), where:
\begin{itemize}
	\item The set of valid state is $Q = (\textit{listed-values} = \{x\in \mathcal{S}\}, \textit{proofs} = \{(p \in \Pi, \widehat{\mathcal{S}} \subseteq \mathcal{S}, \mathsf{P}\in \mathcal{P}_{\mathcal{L}_{\widehat{\mathcal{S}}}})\})$, where \textit{listed-values} is a subset of $\mathcal{S}$ and \textit{proofs} is a set of tuples. Each tuple in \textit{proofs} consists of a proof associated with the set it applies to and to the identifier of the process that issued the proof;
	\item The set of valid initial states is $Q_0 = (\emptyset, \emptyset)$, the state where the \textit{listed-values} and the \textit{proofs} arrays are empty;
	\item The set of possible operation is $O = \{$APPEND($x$), PROVE($y)$, READ()$\}$, with $x,y \in \mathcal{S}$;
	\item The set of possible responses is $R = \bigg\{$True, False, $(\widehat{\mathcal{S}} \subseteq \mathcal{S}, \mathsf{P} \in \mathcal{P}_{\mathcal{L}_{\widehat{\mathcal{S}}}}), \{(p \in \Pi, \widehat{\mathcal{S}}' \subseteq \mathcal{S}, \mathsf{P}'\in \mathcal{P}_{\mathcal{L}_{\widehat{\mathcal{S}}}})\} \bigg\}$, where True is the response to a successful APPEND operation, $(\widehat{\mathcal{S}}, \mathsf{P})$ is the response to a successful PROVE operation, $\{(p, \widehat{\mathcal{S}}', \mathsf{P}')\}$ is the response to a READ operation, and False is the response to a failed operation; and
	\item The transition function is $\Delta$. The PROOF-LIST object type supports $5$ possible transitions. %$o$ represent the invoked operation, $p$ is the process invoking the operation, $q_o$ is the original state of the object, $q_r$ is the resulting state of the object, and $R$ is the response of the object after the invocation.
	We define the $5$ possible transitions of $\Delta$ in \Cref{fig:transition function}.
\end{itemize}

The first transition of the $\Delta$ function models a valid APPEND invocation, a value $y \in \mathcal{S}$ is added to the \textit{listed-values} array by a process in the managers' set $\Pi_M$.
The second transition of the $\Delta$ function represents a failed APPEND invocation. Either the process $p_i$ that invokes this function is not authorized to modify the \textit{listed-values} array, i.e., $p_i \notin \Pi_M$, or the value it tries to append is invalid, i.e., $y \notin \mathcal{S}$.
The third transition of the $\Delta$ function captures a valid PROVE operation, where a valid proof is added to the \textit{proofs} array. The function $\mathsf{C}$ will be used to express the anti-flickering property of the \blist implementation. It is a boolean function that outputs either $0$ or $1$.%The statement proven is determined by the language $\mathcal{L}_{\widehat{\mathcal{A}}}$, where $\widehat{\mathcal{A}}$ is a subset of the values added to the \textit{listed-values} array. %The PROOF-LIST object type does not define how $\widehat{\mathcal{A}}$ must be chosen. This specification will be given for each specific instantiation of this object.
The fourth transition of the $\Delta$ function represents an invalid PROVE invocation. Either the proof is invalid, or the set on which the proof is issued is not a subset of the \textit{listed-values} array.
Finally, the fifth transition represents a READ operation. It returns the \textit{proofs} array and does not modify the object's state.

The language $\mathcal{L}_{\mathcal{A}}$ does not directly depend on the \textit{listed-values} array. Hence, the validity of a PROVE operation will depend on the choice of the set ${\mathcal{A}}$.
%It is important to remark that the statement $x \in \mathcal{L}_{\widehat{\mathcal{A}}}$ can be proven explicitly or using ZKP. The validity of the proof is the same in both cases. Therefore, the synchronization power of the shared object does not depend on the proof technique used. For completeness, we will propose implementations of the \wlist and the \blist object types using ZKP setups.
\section{The consensus number of the \wlist object}\label{sec:wlist}
This section provides an \wlist object specification based on the PROOF-LIST object. The specification is then used to analyze the consensus number of the object type.

We provide a specification of the \wlist object defined as a PROOF-LIST object, where $\mathsf{C}(y, \widehat{\mathcal{S}})=1$ and
\begin{equation}\label{eq:wlistlanguage}
	\forall y \in \mathcal{S}, y \in \mathcal{L}_{{\mathcal{A}}} \Leftrightarrow ({{\mathcal{A}}} \subseteq \mathcal{S} \wedge y \in {{\mathcal{A}}}).
\end{equation}
In other words, $y$ belongs to a set ${\mathcal{A}}$. Using the third transition of the $\Delta$ function, we can see that ${{\mathcal{A}}}$ should also be a subset of the \textit{listed-values} array. Hence, this specification supports proofs of set-membership in \textit{listed-values}. A PROOF-LIST object defined for such language follows the specification of the \wlist. To support this statement, we provide an implementation of the object.
%Furthermore, we see from the PROOF-LIST definition that the \textit{listed-values} array is an append-only array. If we see the \wlist object type as an authorization list, no authorized user can be ''de-authorized''. Therefore, a proof on an outdated version of the \textit{listed-values} array is still valid. We will use this statement to implement an \wlist object in a wait-free manner, using only Atomic Snapshot objects. The proof that such instantiation is given in \cref{th:wlist}

To implement the \wlist object, Algorithm~\ref{fig: SM-ACCUMULATOR} uses two Atomic Snapshot objects. The first one represents the \textit{listed-values} array, and the second represents the \textit{proofs} array. These objects are arrays of $N$ entries. Furthermore, we use a function ''Proof'' that on input of a set $\mathcal{S}$ and an element $y$ outputs a proof that $y \in $ \textit{listed-values}. This function is used as a black box, and can either output an explicit proof---an explicit proof can be the tuple $(y, \mathcal{A})$, where $\mathcal{A}\subseteq $ \textit{listed-values}---or a Zero Knowledge Proof.

\begin{algorithm}
  \begin{pchstack}[center]
\resizebox{14cm}{!}{
\pseudocode[mode=text,codesize=\scriptsize]{
	\textbf{Shared variables} \\
	\pcind AS-LV $\leftarrow$ $N$-dimensions Atomic-Snapshot object, initially  $\{\emptyset\}^N$; \\
	\pcind AS-PROOF $\leftarrow$ $N$-dimensions Atomic-Snapshot object, initially  $\{\emptyset\}^N$; \\
	\textbf{Operation} APPEND($v$) \textbf{is}\\
	\pcln \textbf{If} $(v \in \mathcal{S} ) \land (p \in \Pi_M)$ \textbf{then}\\
	\pcln \pcind local-values $\leftarrow$ AS-LV.Snapshot()$[p]$;\\
	\pcln \pcind AS-LV.Update(local-values $\cup\ v$, $p$);\\
	\pcln \pcind \textbf{Return} true;\\
	\pcln \textbf{Else return} false; \\
	\textbf{Operation} READ() \textbf{is}\\
	\pcln \textbf{Return} AS-PROOF.Snapshot();
	}
\pseudocode[lnstart=6,mode=text,codesize=\scriptsize]{
	\textbf{Operation} PROVE($v$) \textbf{is} \\
	\pcln \textbf{If} $p \notin \Pi_V$ \textbf{then}\\
	\pcln \pcind \textbf{Return} false;\\
	\pcln $\mathcal{A}$ $\leftarrow$ AS-LV.Snapshot();\\
	\pcln \textbf{If} $v \in$ ${\mathcal{A}}$ \textbf{then}\\
	\pcln \pcind $\pi_{\mathrm{set-memb}}$ $\leftarrow$ $\mathsf{Proof}(v\in \mathcal{A})$;\\
	%\pcln \pcind $\pi_{\mathrm{set-memb}}$ $\leftarrow$ Acc$_{\mathrm{set-memb}}$.Prove($v$);\\
	\pcln \pcind proofs $\leftarrow$ AS-PROOF.Snapshot()[$p$];\\
	\pcln \pcind AS-PROOF.Update(proofs $\cup$ ($p, {\mathcal{A}}, \pi_{\mathrm{set-memb}}$), $p$);\\
	\pcln \pcind \textbf{Return} $({\mathcal{A}}, \pi_{\mathrm{set-memb}}$);\\
	\pcln \textbf{Else return} false.
}
}
\end{pchstack}
%\Description{Implementation of an \wlist object using Atomic-Snapshot objects}
\caption{Implementation of an \wlist object using Atomic-Snapshot objects}
\label{fig: SM-ACCUMULATOR}
\end{algorithm}
\begin{theorem}\label{th:wlist}
\Cref{fig: SM-ACCUMULATOR} wait-free implements an \wlist object.
\end{theorem}
\begin{proof}
Let us fix an execution $E$ of the algorithm presented in \cref{fig: SM-ACCUMULATOR}. Each invocation is a sequence of a finite number of local operations and Atomic-Snapshot accesses. Because the Atomic Snapshot primitive can be wait-free implemented in the read-write shared memory model, each correct process terminates each invocation in a finite number of its own steps.

Let $H$ be the history of the execution $E$. We define $\bar{H}$, the completed history of $H$. Any invocation in $H$ can be completed in $\bar{H}$. We give the completed history $\bar{H}$ of $H$:
\begin{itemize}
	\item Any invocation of the APPEND operation that did not reach line 3 can be completed with the line ''\textbf{Return} false'';
	\item Any invocation of the PROVE operation that did not reach line 13 can be completed with the line ''\textbf{Return} false'';
	\item Any invocation of the APPEND operation that reached line 3 can be completed with line 4; and
	\item Any invocation of the PROVE operation that reached line 13 can be completed with line 14.
\end{itemize}
The linearization points of the APPEND, PROVE and READ operations are respectively line $3$, line 13 and line 6.
For convenience, We call any operation in $\bar{H}$ that returns ''false'' an invalid operation.
We verify that each operation in $\bar{H}$ respects the specification:
\begin{itemize}
	\item Any operation in $\bar{H}$ run by a process $p$ that is invalid is an operation that only modifies the internal state of $p$ and that was invoked by a faulty process or that was invoked by a process without the write to invoke the operation. Therefore, these invalid operations do not impact the validity and the progress properties of the \wlist object.
	\item If an APPEND operation invoked by a process $p$ in $\bar{H}$ returns ''true'', it implies that $p$ reached line $3$. Therefore $p$ appended a value $v$ to the array \textit{listed-values} at the index $p$. Process $p$ is the only process able to write at this index. Because the Update operation is atomic, and because $p$ is the only process able to write in AS-LV$[p]$, the \textit{listed-values} array append-only property is preserved. Furthermore, the element added to \textit{listed-value} belongs to the set $\mathcal{S}$, and the process that appends the value belongs to the set of managers $\Pi_M$. Therefore, any invocation of the APPEND operation in $\bar{H}$ that returns ''true'' fulfills the APPEND validity property. Hence, any APPEND invocation in $\bar{H}$ follows the \wlist specification.
	\item If an invocation of the PROVE operation by a process $p$ in $\bar{H}$ returns $({\mathcal{A}}, \pi$), then $p \in \Pi_V$ reached line $13$. Therefore, $p$ appended a proof $\pi$ to the \textit{proofs} array at the index $p$, and the proof is a valid proof that $v \in \mathcal{A}$. Process $p$ is the only process allowed to modify the \textit{proofs} array at this index. There is no concurrency on the write operation. Furthermore, the set ${\mathcal{A}}$, is a subset of the AS-LV array (line $9$). Because the only way to add an element to the AS-LV array is via an APPEND operation, because we consider the linearization point of the PROVE operation to be at line $13$, the PROVE validity property is ensured. The progress property is ensured thanks to the atomicity of the Atomic Snapshot object. If some process executes line $3$ of the APPEND operation at time $t_1$, then any correct process that reaches line $8$ of the PROVE($x$) operation at time $t_2>t_1$ will be valid. Hence, any PROVE invocation in $\bar{H}$ follows the \wlist specification.
	\item A READ operation always returns the values of the AS-PROOF array that were linearized before the execution of line $6$, thanks to the atomicity of the Atomic Snapshot object. Furthermore, the returned value is always a set of successful PROVE operations (AS-PROOF). This set ois compounded of proofs associated to the name of the process that invoked the operation. Therefore, the READ validity property is ensured. Hence, any READ invocation in $\bar{H}$ follows the \wlist specification.
\end{itemize}
All operations in $\bar{H}$ follow the \wlist specification. Thus, $\bar{H}$ is a legal history of the \wlist object type, and $H$ is linearizable. To conclude, the algorithm presented in \cref{fig: SM-ACCUMULATOR} is a wait-free implementation of the \wlist object type.
\end{proof}

\begin{corollary}
The consensus number of the \wlist object type is $1$.
\end{corollary}
\section{The consensus number of the \blist object}\label{sec:blist}
In the following, we propose two wait-free implementations establishing the consensus number of the \blist object type. In this section and in the following, we refer to a \blist with $|\Pi_V|=k$ as a $k$-\blist object. This analysis of this parameter $k$ is the core of the study conducted here. Because it is a statically defined parameter, the knowledge of this parameter can improve efficiency of \blist implementation by reducing the number of processes that need to synchronize in order to conduct a proof.
%Revoir un peu le paragraphe précédent.
\subsection{Lower bound}
\Cref{fig: SnM-ACCUMULATOR lower} presents an implementation of a $k$-consensus object using a $k$-\blist object with $\Pi_M = \Pi_V = \Pi$, and $|\Pi| = k$.
It uses an Atomic Snapshot object, AS-LIST, to allow processes to propose values. AS-LIST serves as a helping mechanism \cite{help}. %All the processes can propose a value in AS-LIST.
In addition, the algorithm uses the progress and the anti-flickering properties of the PROVE operation of the $k$-\blist to enforce the $k$-consensus agreement property. The PROPOSE operation operates as follows. First, a process $p$ tries to prove that the element $0$ is not revoked by invoking PROVE($0$). Then, if the previous operation succeeds, $p$ revokes the element $0$ by invoking APPEND($0$). Then, $p$ waits for the APPEND to be effective. This verification is done by invoking multiple PROVE operations until one is invalid. This behavior is ensured by the progress property of the $k$-\blist object. Once the progress has occurred, $p$ is sure that no other process will be able to invoke a valid PROVE($0$) operation. Hence, $p$ is sure that the set returned by the READ operation can no longer grow. Indeed, the READ operation returns the set of valid PROVE operation that occurred prior to its invocation. If no valid PROVE($0$) operation can be invoked, the set returned by the READ operation is fixed (with regard to the element $0$). Furthermore, all the processes in $\Pi$ share the same view of this set.

Finally, $p$ invokes READ() to obtain the set of processes that invoked a valid PROVE($0$) operation. The response to the READ operation will include all the processes that invoked a valid PROVE operation, and this set will be the same for all the processes in $\Pi$ that invoke the PROPOSE operation. Therefore, up to line $7$, the algorithm solved the set-consensus problem. To solve  consensus, we use an additional deterministic function $f_i : \Pi^i \rightarrow \Pi$, which takes as input any set of size $i$ and outputs a single value from this set.

To simplify the representation of the algorithm, we also use the $\mathsf{separator}()$ function, which, on input of a set of proofs $(\{(p \in \Pi, \{\widehat{\mathcal{S}} \subseteq \mathcal{S}, \mathsf{P} \in \mathcal{P}_{\mathcal{L}_\mathcal{S}})\})$, outputs \textit{processes}, the set of processes which conducted the proofs, i.e. the first component of each tuple.
\begin{algorithm}
\begin{pchstack}[center,space=1em]
\pseudocode[mode=text,codesize=\scriptsize]{
	\textbf{Shared variables} \\
	\pcind $k$-dlist $\leftarrow$ $k$-\blist object; \\
	\pcind AS-LIST $\leftarrow$ Atomic Snapshot object, initially $\{\emptyset\}^k$ \\
	\textbf{Operation} PROPOSE($v$) \textbf{is} \\
	\pcln AS-LIST.update($v, p$); \\
	\pcln $k$-dlist.PROVE($0$);
}
\pseudocode[lnstart=2,mode=text,codesize=\scriptsize]{
	\pcln $k$-dlist.APPEND($0$);\\
	\pcln \textbf{Do}\\
	\pcln \pcind ret $\leftarrow$ $k$-dlist.PROVE($0$);\\
	\pcln \textbf{Until} (ret $\neq$ false);\\
	\pcln \textit{processes} $\leftarrow$ $\mathsf{separator}$($k$-dlist.READ());\\
	\pcln \textbf{Return} AS-LIST.Snapshot()[$f_{|\textit{processes}|}(\textit{processes})$].
}
\end{pchstack}
%\Description{Implementation of a $k$-consensus object using one $k$-\blist object and one Atomic Snapshot}
\caption{Implementation of a $k$-consensus object using one $k$-\blist object and one Atomic Snapshot}
\label{fig: SnM-ACCUMULATOR lower}
\end{algorithm}
\begin{theorem}\label{th: SnM-ACCUMULATOR lower}
\Cref{fig: SnM-ACCUMULATOR lower} wait-free implements a $k$-consensus object.
\end{theorem}
\begin{proof}
Let us fix an execution $E$ of the algorithm presented in \Cref{fig: SnM-ACCUMULATOR lower}. The progress property of the $k$-\blist object ensures that the while loop in line $4$ consists of a finite number of iterations---an APPEND($0$) is invoked prior to the loop, hence, the PROVE($0$) operation will eventually be invalid. Each invocation of the PROPOSE operation is a sequence of a finite number of local operations, Atomic Snapshot object accesses and $k$-\blist object accesses which are assumed atomic. Therefore, each process terminates the PROPOSE operation in a finite number of its own steps.
Let $H$ be the history of $E$. We define $\bar{H}$ the completed history of $H$, where an invocation of PROPOSE which did not reach line 8 is completed with a line "return false". Line $8$ is the linearization point of the algorithm.
For convenience, any PROPOSE invocation that returns false is called an failed invocation. Otherwise, it is called a successful invocation.

We now prove that all operations in $\bar{H}$ follow the $k$-consensus specification:
\begin{itemize}
	\item The process $p$ that invoked a failed PROPOSE operation in $\bar{H}$ is faulty---by definition, the process prematurely stopped before line $8$. Therefore, the fact that $p$ cannot decide does not impact the termination nor the agreement properties of the $k$-consensus object.
	\item A successful PROPOSE operation returns AS-LIST.Snapshot()[$f_{|\textit{processes}|}(\textit{processes})$]. Furthermore, a process proposed this value in line $1$. All the processes that invoke PROPOSE conduct an APPEND($0$) operation, and wait for this operation to be effective using the while loop at line $4$ to $6$. Thanks to the anti-flickering property of the $k$-\blist object, when the APPEND operation is effective for one process---i.e. the Progress happens, in other words,a PROVE($0$) operation is invalid---, then it is effective for any other process that would invoke the PROVE($0$) operation. Hence, thanks to the anti-flickering property, when a process obtains an invalid response from the PROPOSE($0$) operation at line $5$, it knows that no other process can invoke a valid PROVE($0$) operation. This implies that the READ operation conducted at line $7$ will return a fix set of processes, and all the processes that reach this line will see the same set. Furthermore, because each process invokes a PROPOSE($0$) before the APPEND($0$) at line $3$, at least one valid PROPOSE($0$) operation was invoked. Therefore, the \textit{processes} set is not empty. Because each process ends up with the same set \textit{processes}, and thanks to the determinism of the function $f_i$, all correct processes output the same value $v$ (Agreement property and non-trivial value). The value $v$ comes from the Atomic Snapshot object, composed of values proposed by authorized processes (Validity property). Hence a successful PROPOSE operation follows the $k$-consensus object specification.
\end{itemize}
All operations in $\bar{H}$ follow the $k$-consensus specification. To conclude, the algorithm presented in \cref{fig: SnM-ACCUMULATOR lower} is a wait-free implementation of the $k$-consensus object type.
\end{proof}
\begin{corollary}
The consensus number of the $k$-\blist object type is at least $k$.
\end{corollary}

\subsection{Upper bound}
This section provides a \blist object specification based on the PROOF-LIST object. The specification is then used to analyze the upper bound on the consensus number of the object type.

We provide an instantiation of the \blist object defined as a PROOF-LIST object, where:
\begin{equation*}\label{eq:blistlanguage}
	\forall y \in \mathcal{S}, y \in \mathcal{L}_{{\mathcal{A}}} \Leftrightarrow ({\mathcal{A}} \subseteq \mathcal{S} \wedge y \notin {\mathcal{A}}).
\end{equation*}
And where :
\[
    \mathsf{C}(y, \widehat{\mathcal{S}})=
\begin{cases}
    1,& \text{if } \forall \mathcal{A}' \in  \widehat{\mathcal{S}},  y \notin \mathcal{A}'  \\
    0,              & \text{otherwise.}
\end{cases}
\]
In other words, the first equation ensures that $y$ does not belong to a set ${\mathcal{A}}$, while the second equation ensures that the object fulfills the anti-flickering property. Hence, this instantiation supports proofs of set-non-membership in \textit{listed-values}.
A PROOF-LIST object defined for such language follows the specification of the \blist. To support this statement, we provide an
implementation of the object.

%Let $|\Pi_V| = k$. Because of the anti-flickering property, each time a process wants to conduct a set-non-membership proof, all $k$ processes in $\Pi_V$ must agree on the most up-to-date version of the \textit{listed-values} array.
To build a $k$-\blist object which can fulfill the anonymity property, it is required to build an efficient helping mechanism that preserves anonymity. It is impossible to disclose directly the value proven without disclosing the user's identity. Therefore, we assume that a process $p$ that invokes the PROVE($v$) operation can deterministically build a cryptographic commitment to the value $v$. Let $C_v$ be the commitment to the value $v$. Then, any process $p' \ne p$ that invokes PROVE($v$) can infer that $C_v$ was built using the value $v$. However, a process that does not invoke PROVE($v$) cannot discover to which value $C_v$ is linked. %This commitment is used in the implementation presented in \Cref{fig: SnM-ACCUMULATOR higher} to maintain the anti-flickering property.
If the targeted application does not require the user's anonymity, it is possible to use the plaintext $v$ as the helping value.

\Cref{fig: SnM-ACCUMULATOR higher} presents an implementation of a $k$-\blist object using $k$-consensus objects and Atomic Snapshots. The APPEND and the READ operations are analogous to those of \Cref{fig: SM-ACCUMULATOR}.

On the other hand, the PROVE operation must implement the anti-flickering property. To this end, a set of $k$-consensus objects and a helping mechanism based on commitments are used.

When a process invokes the PROVE($v$) operation, it publishes $C_v$, the cryptographic commitment to $v$, using an atomic snapshot object. This commitment is published along with a timestamp \cite{lamport78} defined as follow. A local timestamp $(p, c)$ is constituted of a process identifier $p$ and a local counter value $c$. The counter $c$ is always incremented before being reused. Therefore, each timestamp is unique. Furthermore, we build the strict total order relation $\mathcal{R}$ such that $(p,c)\mathcal{R}(p',c') \Leftrightarrow  (c<c')\lor\left((c=c')\land(p<p')\right)$. The timestamp is used in coordination with the helping value $C_v$ to ensure termination. A process $p$ that invokes the PROVE($v$) operation must parse all the values proposed by the other processes. If a PROVE($v'$) operation was invoked by a process $p'$ earlier than the one invoked by $p$---under the relation $\mathcal{R}$---, then $p$ must affect a set ''val'' for the PROVE operation of $p'$ via the consensus object. The set ''val'' is obtained by reading the AS-LV object. The AS-LV object is append-only---no operation removes elements from the object.  Furthermore, the sets ''val'' are attributed via the consensus object. Therefore, this mechanism ensures that the sets on which the PROVE operations are applied always grow.

Furthermore, processes sequentially parse the CONS-ARR using the counter$_p$ variable. This behavior, in collaboration with the properties of the consensus, ensures that all the process see the same tuples (winner, val) in the same order.

Finally, if a process $p$ observes that a PROVE operation conducted by a process $p' \ne p$ is associated to a commitment $C_v$ equivalent to the one proposed by $p$, then $p$ produces the proof of set-non-membership relative to $v$ and the set ''val'' affected to $p'$ in its name. We consider that a valid PROVE operation is linearized when this proof of set-non-membership is added to AS-PROOF in line $19$.
 Hence, when $p$ produces its own proof---or if another process produces the proof in its name---it is sure that all the PROVE operations that are relative to $v$ and that have a lower index in CONS-ARR compared to its own are already published in the AS-PROOF Atomic Snapshot object. Therefore, the anti-flickering property is ensured.  Indeed, because the affected sets ''val'' are always growing and because of the total order induced by the CONS-ARR array, if $p$ reaches line $25$, it previously added a proof to AS-PROOF in the name of each process $p'\ne p$ that invoked a PROVE($v$) operation and that was attributed a set at a lower index than $p$ in CONS-ARR. Hence, the operation of $p'$ was linearized prior to the operation of $p$. %Thus, ensuring the anti-flickering property

A PROVE operation can always be identified by its published timestamp. Furthermore, when a proof is added to the AS-PROOF object, it is always added to the index counter$_{p_w}$. Therefore, if multiple processes execute line $19$ for the PROVE operation labeled counter$_{p_w}$, the AS-PROOF object will only register a unique value.

%be reached on the set of values considered to ensure the anti-flickering property. To this end, $k$-consensus objects are used. When a process needs to conduct a proof, it communicates its will to do so by publishing

Furthermore, we use a function ''Proof'' that on input of a set $\mathcal{S}$ and an element $x$ outputs a proof that $x \notin \mathcal{S}$. This function is used as a black box, and can either output an explicit proof---an explicit proof can be the tuple $(x, \mathcal{S})$---, or a Zero Knowledge Proof.
\begin{algorithm}
  \begin{pchstack}[center,space=1em]
    \resizebox{14cm}{!}{
      \pseudocode[mode=text,codesize=\scriptsize]{
	\textbf{Shared variables} \\
	\pcind AS-LV $\leftarrow$ $N$-dimensions Atomic-Snapshot object, initially  $\{\emptyset\}^N$; \\
	\pcind AS-Queue $\leftarrow$ $N$-dimensions Atomic-Snapshot object, initially  $\{\emptyset\}^N$;\\
	\pcind CONS-ARR$_p$ $\leftarrow$ an array of $k$-consensus objects of size $l>0$;\\
	\pcind AS-PROOF $\leftarrow$ $l$-dimensions Atomic-Snapshot object, initially  $\{\emptyset\}^l$; \\
	\textbf{Local variables}\\
	\pcind \textbf{For each} $p \in \Pi_V$ :\\
	\pcind \pcind evaluated$_p$ $\leftarrow$ an array of size $l>0$, initially $\{\emptyset\}^l$;\\
	\pcind \pcind counter$_p$ $\leftarrow$ a positive integer, initially $0$;\\
	%\pcind \pcind similar-proofs-array $\leftarrow$ an array of size $l'>0$, initially $\{\emptyset\}^{l'}$\\
	\textbf{Operation} APPEND($v$) \textbf{is}\\
	\pcln \textbf{If} $(v \in \mathcal{S}) \wedge (p \in \Pi_M)$ \textbf{then}\\
	\pcln \pcind local-values $\leftarrow$ AS-LV.Snapshot()$[p]$;\\
	\pcln \pcind AS-LV.UPDATE(local-values $\cup\ v$, $p$);\\
	\pcln \pcind \textbf{Return} true;\\
	\pcln \textbf{Else return} false; \\
	\textbf{Operation} PROVE($v$) \textbf{is} \\
	\pcln \textbf{If} $p \notin \Pi_V$ \textbf{then}\\
	\pcln \pcind \textbf{Return} false;\\
	\pcln \textcolor{black}{$C_v$ $\leftarrow$ Commitment($v$)};
}
\pseudocode[lnstart=8,mode=text,codesize=\scriptsize]{
	\pcln cnt $\leftarrow$ counter$_p$;\\
	\pcln AS-Queue.UPDATE(((cnt, $p$), $C_v$), $p$);\\
	\pcln queue $\leftarrow$ AS-Queue.Snapshot() $\setminus$ evaluated$_p$;\\
	\pcln \textbf{While} (cnt, $p$) $\in$ queue \textbf{do}\\
	\pcln \pcind oldest $\leftarrow$ the smallest clock value in queue under $\mathcal{R}$;\\
	\pcln \pcind prop $\leftarrow$ (oldest, AS-LV.snapshot());\\
	%\pcln \pcind \textbf{If} oldest is still the smallest clock value in queue under $\mathcal{R}$:\\
	\pcln \pcind (winner, val) $\leftarrow$ CONS-ARR[counter$_p$].propose(prop);\\
	\pcln \pcind {((counter$_{p_w}$, $p_w$), $C^*) \leftarrow$ winner;}\\
	\pcln \pcind {\textbf{If}  $C^* = C_v \wedge v \notin$ val \textbf{then}}\\
	\pcln \pcind \pcind {$\pi_{{SNM}}$ $\leftarrow$ Proof($v\notin val$)};\\
	%\pcln \pcind \pcind proofs $\leftarrow$ AS-PROOF.Snapshot()[$p$];\\
	\pcln \pcind \pcind AS-PROOF.Update(($p_w$, val$, \pi_{{SNM}}$, winner), counter$_{p_w}$);\\
	\pcln \pcind evaluated$_p$ $\leftarrow$ evaluated$_p$ $\cup$ winner;\\
	\pcln \pcind queue $\leftarrow$ queue $\setminus$ winner;\\
	\pcln \pcind counter$_p$ $\leftarrow$ counter$_p + 1$;\\
	\pcln \textbf{If} $v \notin$ val \textbf{then}\\
	\pcln \pcind \textbf{Return} (val$, \pi_{\mathrm{SNM}}$);\\
	\pcln \textbf{Else return} false;\\
	\textbf{Operation} READ() \textbf{is}\\
	\pcln \textbf{Return} AS-PROOF.Snapshot();
}
}
\end{pchstack}
%\Description{$k$-\blist object type implementation using $k$-consensus objects and Atomic Snapshot objects.}
\caption{$k$-\blist object type implementation using $k$-consensus objects and Atomic Snapshot objects.}
\label{fig: SnM-ACCUMULATOR higher}
\end{algorithm}
\begin{theorem}\label{th: SnM-ACCUMULATOR higher}
\Cref{fig: SnM-ACCUMULATOR higher} wait-free implements a $k$-\blist object.
\end{theorem}
\begin{proof}
	Let us fix an execution $E$ of the algorithm presented in \Cref{fig: SnM-ACCUMULATOR higher}. The strict order relation $\mathcal{R}$ used to prioritize accesses to the CONS-ARR array implies that each process that enters the while loop in line $12$ will only iterate a finite number of times. Furthermore, we assume that $k$-consensus objects and atomic-snapshot objects are atomic. Therefore, each process returns from a PROVE, an APPEND, or a READ operation in a finite number of its own steps.

	Let $H$ be the history of $E$. We define $\bar{H}$, the completed history of $H$. We associate a specific response with all pending invocations in $H$. The associated responses are:
	\begin{itemize}
		\item Any invocation of the APPEND operation that did not reach line 3 can be completed with the line ''\textbf{Return} false''.
		\item Any invocation of the PROVE operation that did not reach line $10$ can be completed with the line ''\textbf{Return} false''.
		\item Any pending invocation of the PROVE operation by the process $p$ that reached line $10$ is completed with the line ''\textbf{Return} (val$, \pi_{SNM}$);'' if ($p$, value$, \pi_{SNM}$, winner) is in the AS-PROOF array, and the value added by process $p$ in line $10$ is ''winner''. Otherwise, the operation is completed with the line ''\textbf{Return} false''.
		\item Any pending invocation of the APPEND operation that reached line $3$ can be completed with line $4$.
	\end{itemize}
	The linearization point of the APPEND and READ operations are respectively at line $3$ and $26$. Let us consider a valid PROVE operation invoked by a process $p$ that is attributed a tuple (winner, val) at the index counter$_{p_w}$ of the CONS-ARR array. We say this operation is linearized when the first AS-PROOF.Update labeled with counter$_{p_w}$ in line $19$ is executed by any process.

	 For convenience, we call operations that return false invalid operations. The consensus objects in CONS-ARR are accessed at most once by each process. There are only $k=|\Pi_V|$ processes allowed to access these objects. Therefore, the $k$-consensus objects in the array always return a value different from $\emptyset$. We now prove that all operations in $\bar{H}$ follow the \blist specification:
	 \begin{itemize}
	 	\item An invalid APPEND operation in $\bar{H}$ only modifies the internal state of the process. This operation does not modify the state of the shared object. It is either invoked by an unauthorized process which fails in line $1$, or by a faulty process. This operation follows the specification;
	 	\item An invalid PROVE operation in $\bar{H}$ is an operation that returns false in line $7$ or $25$. In the first case, the process was not authorized to propose a proof. In the second case, the value $v$ used by the process is already inside the set ''val'' the process was attributed by the consensus in line $15$. This set is produced from the values added to the AS-LV object. This object begins as an empty set, and values inside this set can only be added using the APPEND operation. Therefore, the PROVE validity property is ensured.
	 	\item If an invocation of the APPEND operation in $\bar{H}$ returns true, it implies that process $p$ appended a value $v$ to the  \textit{listed-values} array, at the index $p$ at line $3$. Because the WRITE operation is atomic, and because $p$ is the only process able to write in AS-ACC[$p$], the \textit{listed-values} array append-only property is preserved. Hence a successful APPEND operation follows the specification.
	 	\item If an invocation of the PROVE operation in $\bar{H}$ returns True, it implies that: 1) process $p$ was attributed a $k$-consensual set ''val'' on line $15$, 2) from line $17$, $v \notin$ val, and 3) a proof that $v \notin$ ''val'' was added to the AS-PROOF object, either by $p$ or by another process performing the helping mechanism. First, to prove the progress property, we assume a history where first, a process $p'$ obtains a positive response from an APPEND($v$) operation. Afterward, a process $p$ invokes a PROVE($v$) operation. Therefore, the value $v$ will already be included in the AS-LV object at this time because $p'$ received a positive response from its invocation. Any process that executes the line $14$ of the PROVE operation after the invocation of $p$ will propose a set where $v$ is included. Therefore, the set ''val'' that will be affected to $p$ by the consensus on line $15$ will include $v$. The PROVE($v$) operation invoked by $p$ will be invalid. The progress property is ensured.

	 	Second, the anti-flickering property is ensured by the helping mechanism and the $k$-consensus objects used from line $10$ to $22$. The processes in $\Pi_V$ that invoke the PROVE($v$) operation will sequentially attribute a set ''val'' to each proving process, using the set of $k$-consensus objects. Furthermore, this sequential attribution takes into account the evolution of the AS-LV object. Therefore, the set associated with the object CONS-ARR[$i-1$] is always included in the set associated with the object CONS-ARR[$i$].

	 	Furthermore, the CONS-ARR array is browsed sequentially by each process invoking the PROVE operation. Therefore, if a process $p$ that invokes a PROVE($v$) operation with a timestamp $t$, and this invocation is not valid in the end, $p$ will nonetheless linearize all the PROVE$(v$) operations that have a lower timestamp than $t$ before returning from the operation. Hence, all the valid PROVE($v$) operations will be linearized before the response of $p$'s invocation, and any invocation of a PROVE($v$) operation that occurs after the response of $p$'s invocation will fail.

	 	Third, the PROVE validity property directly follows from point (2) and the anti-flickering property. Hence a successful PROVE operation follows the specification.
	 \item A READ operation always returns, and thanks to the atomicity of the Atomic Snapshot object, it always returns the most up-to-date version of the AS-PROOF array.
	 \end{itemize}
	 All operations in $\bar{H}$ follow the $k$-\blist specification. Therefore the algorithm presented in \Cref{fig: SnM-ACCUMULATOR higher} is a wait-free implementation of the $k$-\blist object type.
       \end{proof}
The following corollary follows from  \Cref{th: SnM-ACCUMULATOR lower} and \Cref{th: SnM-ACCUMULATOR higher}.
\begin{corollary}
The $k$-\blist object type has consensus number $k$.
\end{corollary}

\section{Discussion}\label{sec:discussion}
This section presents several applications where the \wlist and the $k$-\blist can be used to determine consensus number of more elaborate objects.
More importantly, the analysis of the consensus of these use cases makes it possible to determine if actual implementations
achieve optimal efficiency in terms of synchronization. If not, we use the knowledge of the consensus number of the \wlist and \blist objects to
give intuitions on how to build more practical implementations. More precisely, the fact that consensus numbers of
\wlist and \blist objects are (in most cases) smaller than $n$ implies that most implementations can reduce the number of processes that need to synchronize in order to implement such distributed objects.
The liveness of many consensus protocols is only ensured when the network reaches a synchronous period. Therefore, reducing the number of processes that need to synchronize can increase the system's probability of reaching such synchronous periods.
Thus, it can increase the effectiveness of such protocols.

\subsection{Revocation of a verifiable credential}
We begin by analyzing Sovrin's Verifiable-Credential revocation method using the \blist object~\cite{Sovrin}. Sovrin is a privacy-preserving Distributed Identity Management System (DIMS). In this system, users own credentials issued by entities called issuers. A user can employ one such credential to prove to a verifier they have certain characteristics. An issuer may want to revoke a user's credential prematurely. To do so, the issuer maintains an append-only list of revoked credentials. When a user wants to prove that their credential is valid, they must provide to the verifier a valid ZKP of set-non-membership proving that their credential is not revoked, i.e. not in the \blist. In this application, the set of managers $\Pi_M$ consists solely of the credential's issuer.
Hence, the proof concerns solely the verifier and the user. %, which are the only elements of the set $\Pi_M$.
The way Sovrin implements this verification interaction is by creating an ad-hoc peer-to-peer consensus instance between the user and the verifier for each interaction. Even if the resulting \blist has consensus number $2$, Sovrin implements the APPEND operation using an SWMR stored on a blockchain-backed ledger (which requires synchronizing the $N$ processes of the system). Our results suggest instead that Sovrin's revocation mechanism could be implemented without a blockchain by only using pairwise consensus. % Hence Sovrin's revocation mechanism uses at least an object with a consensus number of $n$. the consensus number of this system is $|\Pi|=n$ and is therefore sub-optimal.

\subsection{The Anonymous Asset Transfer object}
The anonymous asset transfer object is another application of the \blist and the \wlist objects. As described in \Cref{sec:aat}, it is possible to use these objects to implement the asset transfer object described in~\cite{Guerraoui}.
Our work generalizes the result by Guerroui et al.~\cite{Guerraoui}. Guerraoui et al. show that a joint account has consensus number $k$ where $k$ is the number of agents that can withdraw from the account. We can easily prove this result by observing that withdrawing from a joint account requires a denylist to record the already spent coins. Nevertheless, our ZKP capable construction makes it possible to show that an asset transfer object where the user is anonymous, and its transactions are unlinkable also has consensus number $k$, where $k$ is the number of processes among which the user is anonymous. The two main implementations of Anonymous Asset Transfer, ZeroCash and Monero \cite{monero, zcash}, use a blockchain as their main double spending prevention mechanism. While the former provides anonymity on the whole network, the second only provides anonymity among a subset of the processes involved in the system. Hence, this second implementation could reduce its synchronization requirements accordingly.

Furthermore, our implementation uses \wlist as \textit{ex-nihilo}-coin-creation prevention mechanism.
Hence, both security properties of an anonymous asset transfer object can be enforced separately.
Using the fact that \wlists have consensus number $1$, this result implies that this part of the protocol could be handled more efficiently.
For example, the most space and resource-intensive part of the ZeroCash protocol is the \textit{ex-nihilo}-coin-creation prevention mechanism.
This part of the protocol is implemented in a synchronous way, i.e., all processes synchronize to conduct it, which is, as we demonstrated, sub-optimal.
Therefore, this analysis may lead to a more efficient ZeroCash-like asset transfer protocol implementation.

\subsection{Distributed e-vote systems}
Finally, another direct application of the \blist object is the blind-signature-based e-vote system with consensus number $k$, $k$ being the number of voting servers, which we present in \Cref{sec:evote}. Most distributed implementations of such systems also use blockchains, whereas only a subset of the processes involved actually require synchronization.
\section{Related Works}
\paragraph*{Bitcoin and blockchain}
%Nakamoto popularized the distributed consensus with its famous asset-transfer protocol, Bitcoin \cite{BTC}. Bitcoin uses a consensus algorithm to maintain a list of transactions. This list is shared among all the processes of the network. The hype around Bitcoin conducted numerous actors to suggest new blockchain-based asset-transfer protocols. The most famous one is, without a doubt, the Ethereum blockchain \cite{ETH}. This alternative blockchain proposes a Deterministic Distributed State Machine implementation along with the asset transfer protocol. The coherence of the state machine is maintained using the underlying consensus algorithm.

Even though distributed consensus algorithms were already largely studied \cite{pbft, paxos, PRpbft, rbft, Bracha}, the rise of Ethereum---and the possibilities offered by its versatile smart contracts---led to new ideas to decentralized already known applications. Among those, e-vote and DIMS~\cite{Sovrin} are two examples.

Blockchains increased the interest in distributed versions of already existing algorithms. However, these systems are usually developed with little concern for the underlying theoretical basis they rely on. A great example is trustless money transfer protocols or crypto money. The underlying distributed asset-transfer object was never studied until recently. A theoretical study proved that a secure asset-transfer protocol does not need synchronicity between network nodes \cite{Guerraoui}. Prior to this work, all proposed schemes used a consensus protocol, which cannot be deterministically implemented in an asynchronous network \cite{FLP}. The result is that many existing protocols could be replaced by more efficient, Reliable Broadcast \cite{Bracha} based algorithms. This work leads to more efficient implementation proposal for money transfer protocol \cite{Auvolat}.
Alpos et al. then extended this study to the Ethereum ERC20 smart contracts \cite{Zanolini}. This last paper focuses on the asset-transfer capability of smart contracts. Furthermore, the object described has a dynamic consensus number, which depends on the processes authorized to transfer money from a given account. Furthermore, this work and the one from Guerraoui et al. \cite{Guerraoui} both analyze a specific object that is not meant to be used to find the consensus number of other applications. In contrast, our work aims to be used as a generic tool to find the consensus number of numerous systems.
%However, many implementations

\paragraph*{E-vote}
An excellent example of the usage of \blist is to implement blind signatures-based e-vote systems \cite{chaum-blind-sig}.
 A blind signature is a digital signature where the issuer can sign a message without knowing its content. Some issuer signs a cryptographic commitment---a cryptographic scheme where Alice hides a value while being bound to it \cite{Pedersen}---to a message produced by a user. Hence, the issuer does not know the actual message signed. The user can then un-commit the message and present the signature on the plain-text message to a verifier. The verifier then adds this message to a \blist. A signature present in the \blist is no longer valid. Such signatures are used in some e-vote systems \cite{e-vote,FOO-improve}. In this case, the blind signature enables anonymity during the voting operation. This is the e-vote mechanism that we study in this article. They can be implemented using a \blist to restrain a user from voting multiple times. This method is explored in \Cref{sec:evote}

There exists two other way to provide anonymity to the user of an e-vote system. The first one is to use a MixNet \cite{JCJ, mixnet, Civitas}. MixNet is used here to break the correlation between a voter and his vote.
Finally, anonymity can be granted by using homomorphic encryption techniques \cite{Pointcheval-homomorphic, Cramer-homomorphic}.

Each technique has its own advantages and disadvantages, depending on the properties of the specific the e-vote system. We choose to analyze the blind signature-based e-vote system because it is a direct application of the distributed \blist object we formalize in this paper.

\paragraph*{Anonymous Money Transfer}

Blockchains were first implemented to enable trustless money transfer protocols. One of the significant drawbacks of this type of protocol is that it only provides pseudonymity to the user. As a result, transfer and account balances can be inspected by anyone, thus revealing sensitive information about the user. Later developments proposed hiding the user's identity while preventing fraud. The principal guarantees are double-spending prevention---i.e., a coin cannot be transferred twice by the same user---and \textit{ex nihilo} creation prevention---i.e., a user cannot create money. Zcash \cite{zcash} and Monero \cite{monero} are the best representative of anonymous money transfer protocols. The first one uses an \wlist to avoid asset creation and a \blist to forbid double spending, while the second one uses ring signatures. We show in \Cref{sec:aat} that the \blist and \wlist objects can implement an Anonymous Money Transfer object, and thus, define the synchronization requirements of the processes of the system.
%\paragraph*{Self Sovereign identities}

%Self Sovereign Identity \cite{survey-ssi} is a new way of implementing Identity Management Systems (IMS). The concept was first proposed by Allen \cite{ten-principles} in 2014. The idea of this type of IMS is to let the user in control of his identity elements. It is to say that, when some user want to prove a statement about himself, he only need to provide a document that is saved on his device. Whereas in previous paradigms, identity elements were stored on a third party server. To achieve SSI, most system \cite{Sovrin, uport} use signatures, issued by a trusted issuer, and stored in a user-controlled wallet. These signature are called verifiable credentials. Different properties can be given to thes signatures, the most advanced ones in term of user's privacy are the Anonymous Credential, proposed by Chaum in 1985 \cite{chaum} and later improved by Camenish and lysyanskaya \cite{AC-Camenish, AC-Came-BP}.

%In order to implement a fully functionnal SSI framework, the Verifiable Credential is not sufficient. The framework needs to allow the actors of the scheme to share information. Because the philosophy of SSI system is to avoid trusted third parties and centralization points, most proposed implementations use some kind of distributed ledgers. These ledgers are used to share information about actors, to authorize usage of credential by specific parties, and to revoke credentials in case of misusage, or in case of theft. These functionalities are based on distributed \wlist and \blist objects.
\section{Conclusion}
This paper presented the first formal definition of distributed \wlist and \blist object types. These definitions made it possible to analyze their consensus number. This analysis concludes that no consensus is required to implement an \wlist object. %In a message passing setup, this result implies that a \wlist object can be implemented using the efficient Reliable Broadcast primitive.
On the other hand, with a \blist object, all the processes that can propose a set-non-membership proof must synchronize, which makes the implementation of a \blist more resource intensive.

The definition of \wlist and \blist as distributed objects made it possible to thoroughly study other distributed objects that can use \wlist and \blist as building blocks.
%These results have direct implications for numerous systems.
For example, we discussed authorization lists and revocation lists in the context of the Sovrin DIMS. We also provided several additional examples in the Appendix. In particular, we show in \Cref{sec:aat} that an association of \blist and \wlist objects can implement an anonymous asset transfer protocol and that this implementation is optimal in terms of synchronization power. This result can also be generalized to any asset transfer protocol, where the processes act as proxies for the wallet owners. In this case, synchronization is only required between the processes that can potentially transfer money on behalf of a given wallet owner.

\newpage
\bibliographystyle{unsrt}
\bibliography{main}
\appendix
\section{Variations on the \textit{listed-values} array}
\newcommand{\kar}{\ensuremath{k_{\textsc{AR}}}}
\label{sec:variations}
In the previous sections, we assumed the \textit{listed-values} array was append-only. Some use cases might need to use a different configuration for this array.
In this section, we want to explore the case where the \textit{listed-values} array is no longer append-only.

\paragraph*{One-process only}

We will first explore a limited scenario where the processes can only remove the values they wrote themselves. In this case, there are no conflicts on the append and remove operations. The \textit{listed-values} array can be seen as an array of $|\Pi_V|$ values. A process $p_i$ can write the $i$-th index of the \textit{listed-values} array. It is the only process that modifies this array. Therefore, there are no conflicts upon writing. We would need to add a REMOVE operation to the \wlist and \blist object. Because of this REMOVE operation, the \wlist could act as a \blist. Indeed, let us assume the managers adds all elements of the universe of the possible identifiers to the \wlist in the first place. Then, this \wlist can implement a \blist object, where the REMOVE operation of the \wlist is equivalent to the APPEND operation of the \blist. Hence, the \wlist object would need an anti-flickering property to prevent concurrent PROVE operations from yielding conflicting results. This implies that an \wlist object implemented with a REMOVE operation is equivalent to a \blist object and has consensus number $k$, where $k$ is the number of processes in $\Pi_V$.

\paragraph*{Multi-process}

The generalization of the previous single-write-remove
\textit{listed-values} array is a \textit{listed-values} array where
$\kar$ ($\textsc{AR}$ for APPEND/REMOVE) processes can remove a value
appended by process $p_i$. We assume each process $p$ is authorized to
conduct APPEND and REMOVE operations on its ''own''
register. Furthermore, each process $p_i$ has a predefined
authorization set $\mathcal{A}_i \subseteq \Pi_M$, defining which
processes can APPEND or REMOVE on $p_i$'s register. We always have
$p_i \in \mathcal{A}_i$. If $p_j \in \mathcal{A}_i$, then $p_j$ is
allowed to ''overwrite'' (remove) anything $p_i$ wrote. In this case,
all authorized processes need to synchronize in order to write a value
on the \textit{listed-values} array. More precisely, we can highlight
two cases.

The first case is the ''totally shared array'' case, where all
processes share the same $\mathcal{A}_i = \Pi_M$. Any modifications on
the \textit{listed-values} array by one process $p_i$ can be in
competition with any other process $p_j \in \Pi_M$. Therefore, there
must be a total synchronization among all the processes of the managers'
set to modify the \textit{listed-values} array. When such behaviour is
needed, both \wlist and \blist require solving consensus among at
least $|\Pi_M|$ processes to implement the APPEND and REMOVE
operations.
% DAVIDE: I removed this sentence because I think it may be wrong. Given an object O, CN(O)=k means that you can implement consensus among k processes using O; however, here we have an object that has two operations where one APPEND/REMOVE NEEDS consensus among k1 and the other PROVE, needs consensus among k2. Without further information, I think we cannot say whether the CN is min(k1, k2) or max (k1,k2). When such behaviour is needed, both \wlist and \blist object types have a consensus number of at least $|\Pi_M|$, considering the operations APPEND and REMOVE.

The second case is the ''cluster'' case: a subset of processes share a
sub-array, which they can write. In this case, each process in a
given cluster must synchronize before writing (or removing) a
value. The synchronization required is only between this cluster's
$\kar$ authorized process. This corresponds to some extent to a sharded
network~\cite{sharding}. % The overall \textit{listed-values} array can
% be seen as the one-process authorization setup, where each node is
% managed via a $\kar$-consensus algorithm by multiple processes. This
% behaviour should limit the need for a total consensus over the network
% and should improve the overall performance of the network. However, it
% raises new issues, such as committee selection. These issues have been
% studied in the context of sharded blockchains.

\section{Anonymous Asset-Transfer object type}\label{sec:aat}
Decentralized money transfer protocols were popularized by Bitcoin \cite{BTC}. Guerraoui et al. proposed a  theoretical analysis \cite{Guerraoui} that proved that the underlying object, the asset transfer object, has consensus number 1 if each account is owned by a single process. This result implies that the expensive Proof Of Work (POW) leveraged by the Bitcoin implementation is an over-engineered solution in a message-passing setting. A less expensive solution based on the Reliable-Broadcast primitive works as well \cite{Auvolat}. The paper by Guerraoui et al. also studies the case where multiple processes share accounts. In this case, the consensus number of the resulting object is $k$, the maximum number of processes sharing a given account.

These works give a good insight into the problem of asset transfer, but they only study pseudonymous systems, where all transactions can be linked to a single pseudonym. With the growing interest in privacy-enhancing technologies, cryptocurrency communities try to develop anonymous and unlinkable money transfer protocols \cite{zcash, monero}. The subsequent question is to know the consensus number associated with the underlying distributed object. The formalization of the \wlist and the \blist objects presented in this article makes it possible to answer this question. This section is dedicated to this proof.

\subsection{Problem formalization}

\paragraph*{Asset-Transfer object type definition}
The Asset-Transfer object type allows a set of processes to exchange assets via a distributed network.
We reformulate the definition proposed by Guerraoui et al. \cite{Guerraoui} to describe this object:

\begin{definition}\label{def: AT}
The (pseudonymous) Asset-Transfer object type proposes two operations, \emph{TRANSFER} and \emph{BALANCE}. The object type is defined for a set $\Pi$ of processes and a set $\mathcal{W}$ of accounts. An account is defined by the amount of assets it contains at time $t$. Each account is initially attributed an amount of assets equal to $v_0 \in \ZZ^{+*}$. We define a map $\mu: \mathcal{W} \rightarrow \{0, 1\}^{|\Pi|}$ which associates each account to the processes that can invoke \emph{TRANSFER} operations for these wallets. The Asset Transfer object type supports two operations, \emph{TRANSFER} and \emph{BALANCE}. When considering a \emph{TRANSFER}($i,j,v$) operation, $i \in \mathcal{W}$ is called the initiator, $j \in \mathcal{W}$ is called the recipient, and $v \in \NN$ is called the amount transferred. Let $T(i,j)_t$ be the sum of all valid \emph{TRANSFER} operations initiated by process $i$ and received by process $j$ before time $t$. These operations respect three properties:
	\begin{itemize}
		\item (Termination) \emph{TRANSFER} and \emph{BALANCE} operations always return if they are invoked by a correct process.
		\item (\emph{TRANSFER} Validity) The validity of an operation \emph{TRANSFER}($x, y, v$) invoked at time $t$ by a process $p$ is defined in a recursive way. If no \emph{TRANSFER}($x, i, v$), $\forall i \in \mathcal{W}$ was invoked before time $t$, then the operation is valid if $v \le v_0$ and if $p \in \mu(x)$. Otherwise, the operation is valid if $v\le v_0 + \sum_{i \in \mathcal{W}} T(i,x)_t - \sum_{j \in \mathcal{W}} T(x,j)_t $ and if $p \in \mu(x)$.
		\item (\emph{BALANCE} Validity) A \emph{BALANCE} operation invoked at time $t$ is valid if it returns $v_0 + \sum_{i \in \mathcal{W}} T(i,x)_t - \sum_{j \in \mathcal{W}} T(x,j)_t $ for each account $x$.
	\end{itemize}
\end{definition}

The Asset transfer object is believed to necessitate a double-spending-prevention property. This property is captured by the TRANSFER Validity property of \Cref{def: AT}. Indeed, the double-spending-prevention property is defined to avoid ex-nihilo money creation. In a wait-free implementation, a valid transfer operation is atomic. Therefore, double spending is already prevented. A TRANSFER operation takes into account all previous transfers from the same account.

The paper by Guerraoui et al. \cite{Guerraoui} informs us that the consensus number of such an object depends on the map $\mu$. If $\sum_{i\in \{0, \cdots, |\Pi|\}} \mu(w)[i] \le 1, \forall\ w \in \mathcal{W}$, then the consensus number of the object type is $1$. Otherwise, the consensus number is $\max_{w\in\mathcal{W}}(\sum_{i\in \{0, \cdots, |\Pi|\}} \mu(w)[i])$. In other words, the consensus number of such object type is the maximum number of different processes that can invoke a TRANSFER operation on behalf of a given wallet.

\paragraph*{From continuous balances to token-based Asset-Transfer}

The definition proposed by Guerraoui et al. uses a continuous representation of the balance of each account. Implementing anonymous money transfer with such a representation would require a mechanism to hide the transaction amounts~\cite{zcash}. As such a mechanism would not affect the synchronization properties of the anonymous-money-transfer object, we simplify the problem by considering a token-based representation. This means that the algorithm can transfer only tokens of a predefined weight. To move from one representation to the other, we operate a bijection between the finite history of a continuous account-based money representation and a token-based representation.

Let $\mathcal{W}$ be a set of accounts. Let $\mu: \mathcal{W} \rightarrow \{0,1\}^{\Pi}$ be the owner map, and let $D_{(I,V)}$ be a discretization function such that $D_{(I,V)}: \mathcal{S}\subseteq\RR^+ \rightarrow \NN^I$, for some $I\in \NN^*$. $D_{(I,V)}$ is defined as follows:
\begin{align*}
\forall i \in \{0,\cdots, I\},\ \forall x\in \mathcal{S}: & \sum_{i=1}^{I} D_{(I,V)}(x)[i] = x \\
& \text{ where } D_{(I,V)}(x)= \left((D_{(I,V)}(x)[i] = V)\lor(D_{(I,V)}(x)[i] = 0)\right). \\
\end{align*}
We need to define $\mathcal{S}$ in order to make $D_{(I,V)}$ bijective. Let the array $\{InitBal_1, \cdots, InitBal_{|\mathcal{W}|}\}$ be the initial balances of the $|\mathcal{W}|$ accounts, with $InitBal_i \in \RR^+, \forall i\in \{1, \cdots, p\}$. Let us fix an execution E of an Asset Transfer object. Let H be the history of E, and let $\bar{H}$ be the linearization of this history. Let us assume the amounts transferred are in $\NN^{*+}$. Let $T$ be the array of transferred amounts. Let $V$ be the greatest common divisor of all the elements in $T$ and the initial balance array. Let $\textit{TotalTrans}_i$ be the total number of transactions the account receives $i$ in $\bar{H}$. Let $balance(i,j)$ be the balance of the account $i$ after its $j$'th asset reception. Let $I$ be the greatest amount of money possessed by an account in $\bar{H}$ divided by $V$, i.e., $I=\frac{\max_{i \in \{1, \cdots, |\mathcal{W}|\}}\left( max_{j \in \{1,\cdots, \textit{TotalTrans}_i\}} (balance(i,j)) \right)}{V}$. Hence, we can apply $D_{(I,V)}$ to each element of the initial balance array. We can define $\mathcal{S}$ as the infinite set $\{0, V, 2V, 3V, \cdots\}$.  Using such an $\mathcal{S}$, the map $D_{(I,V)}$ is bijective when applied to $\bar{H}$.

Thanks to this bijection, all transactions can be seen as a given amount of ''coins'' transferred. A coin corresponds to a given amount of asset $V$. The amount of coins of each account is represented by an array of size $I$, with values $V$ or $0$. Hence, we have a discretized version of the asset transfer object, and there exists a bijection between the continuous setup and the discretized setup.

%Now, we want to "tokenize" this representation, i.e., we want each coin to be unique. We have arrays of discrete values. We can take the initial balance array and apply $D_{(I,V)}$. Each account balance is now represented by an array $coins = \{V,V, \cdots, 0, 0\}$. Each account has a unique name. We can therefore name each "coin" with a unique tuple $(p,i)$ where $p$ is the name of the owner's account and $i$ is the index of this "coin" in the $coins$ array\footnote{We can also use the hash of these tuples as names. Thanks to the collision resistance property of hash functions, the uniqueness is maintained.}. We call these tuples tokens. Account owners can exchange tokens in the same way as with the continuous representation, except that now the verification function is based on the total number of tokens possessed rather than on the balance (which are ultimately the same values). Because the token version of the AT-object is bijective with the discrete version, this tokenized version is also bijective with the original version of the AT-object.

We use the discrete version of the Asset Transfer object in the following to reason about Anonymous Asset transfers. Specifically, a transfer in the tokenized version for a value of $kV$ consists of $k$ TRANSFER operations, each transferring a token of value $V$.

\paragraph*{Anonymity set}

Let $S$ be a set of actors. We define "anonymity" as the fact that, from the point of view of an observer, $o \notin S$, the action, $v$,  of an actor, $a \in S$, cannot be distinguished from the action of any other actor, $a' \in S$. We call $S$ the anonymity set of $a$ for the action $v$ \cite{Anon-def}.

Implementing Anonymous Asset Transfer requires hiding the association between a token and the account or process that owns it. If a "token owner" transfers tokens from the same account twice, these two transactions can be linked together and are no longer anonymous. Therefore, we assume that the ''token owner'' possesses offline proofs of ownership of tokens. These proofs are associated with shared online elements, allowing other processes to verify the validity of transactions. We call \textit{wallet} the set of offline proofs owned by a specific user. We call the individual who owns this wallet the \textit{wallet owner}. It is important to notice that a wallet owner can own multiple wallets, whereas we assume a wallet is owned by only one owner.
 Furthermore, we assume each process can invoke TRANSFER operations on behalf of multiple wallet owners. Otherwise, a single process, which is in most cases identified by its ip-address or its public key, would be associated with a single wallet. Thus, the wallet would be associated with a unique identifier, and the transactions it would operate could not be anonymous. With the same reasoning, we can assume that a wallet owner can request many processes to invoke a TRANSFER operation on his or her behalf. Otherwise, the setup would not provide "network anonymity", but only "federated anonymity", where the wallet is anonymous among all other wallets connected to this same process. In our model, processes act as proxies.

\paragraph*{The Anonymous Asset-Transfer object type}
We give a new definition of the Asset-Transfer object type that takes anonymity into account. The first difference between a Pseudonymous Asset Transfer object type and an anonymous one is the absence of a BALANCE operation. The wallet owner can compute the balance of its own wallet using a  LOCALBALANCE function that is not part of the distributed object. The TRANSFER operation is also slightly modified. Let us consider a sender that wants to transfer a token $T_O$ to a recipient. The recipient creates a new token $T_R$ with the associated cryptographic offline proofs (in practice, $T_R$ can be created by the sender using the public key of the recipient). Specifically, it associates it with a private key. This private key is known only to the recipient: its knowledge represents, in fact, the possession of the token. Prior to the transfer operation, the recipient sends token $T_R$ to the sender. The sender destroys token $T_O$ and activates token $T_R$. The destruction prevents double spending, and the creation makes it possible to transfer the token to a new owner while hiding the recipient's identity. Furthermore, this process of destruction and creation makes it possible to unlink the usages of what is ultimately a unique token.

Each agent maintains a local wallet that contains the tokens (with the associated offline proofs) owned by the agent. The owner of a wallet $w$ can invoke TRANSFER operations using any of the processes in $\mu(w)$. A transfer carried out from a process $p$ for wallet $w$ is associated with an anonymity set $\mathcal{AS}_p^w$ of size equal to the number of wallets associated with process $p$:  $|\mathcal{AS}_p^w|=\sum_{i\in\mathcal{W}} \mu(i)[p]$. The setup with the maximal anonymity set for each transaction is an Anonymous Asset Transfer object where each wallet can perform a TRANSFER operation from any process: i.e., $\mu(i) = \{1\}^{|\Pi|}, \forall i\in\mathcal{W}$. The token-based Anonymous Asset Transfer object type is defined as follows:

%\footnote{It directly follows from this definition and comments in paragraph ''Asset-Transfer object type definition'' that the consensus number of the Anonymous Asset Transfer object depends on the size of this anonymity set. The following sections will clarify this point.}

%% In our construction, when a TRANSFER($T_O, T_R$) operation is invoked, the token associated with $T_O$ is considered ''destroyed'', and the token $T_R$ is created right after this destruction.

\begin{definition}\label{def: AAT}
The Anonymous Asset Transfer object type supports only one operation: the \emph{TRANSFER} operation. It is defined for a set $\Pi$ of processes and a set $\mathcal{W}$ of wallets. An account is defined by the amount of tokens it controls at time $t$. Each account is initially attributed an amount $v_0$ of tokens. We define a map $\mu: \mathcal{W} \rightarrow \{0, 1\}^{|\Pi|}$ which associates each wallet to the processes that can invoke \emph{TRANSFER} on behalf of these wallets.  When considering a \emph{TRANSFER}($T_O, T_R$) operation, $T_0$ is the cryptographic material of the initiator that proves the existence of a token $T$, and $T_R$ is the cryptographic material produced by the recipient used to create a new token. The \emph{TRANSFER} operation respects three properties:
	\begin{itemize}
		\item (Termination) The \emph{TRANSFER} operation always returns if it is invoked by a correct process.
		\item (\emph{TRANSFER} Validity) A \emph{TRANSFER}($T_O, T_R$) operation invoked at time $t$ is valid if:
			\begin{itemize}
				\item (Existence) The token $T_O$ already existed before the transaction, i.e., either it is one of the tokens initially created, or it has been created during a valid \emph{TRANSFER}($T_O', T_O$) operation invoked at time $t'<t$.
				\item (Double spending prevention) No \emph{TRANSFER}($T_O$, $T_R'$) has been invoked at time $t''<t$.
			\end{itemize}
		\item (Anonymity) A \emph{TRANSFER}($T_O, T_R$) invoked by process $p$ does not reveal information about the owner $w$ and $w'$ of $T_O$ and $T_R$, except from the fact that $w$ belongs to the anonymity set $\mathcal{AS}_p^w$.
	\end{itemize}
\end{definition}

%We define the TRANSFER operation for a $k$-anonymous-Asset-Transfer object type to be an operation where the sender of a transaction selects an anonymity set $S$ of $k$ potential senders and produces a proof that he can spend the required amount (no creation of money), and that this amount has not already been spent (double-spending protection). Because the sender must be anonymous among the $k$ other potential senders, this implies that the token could be spent by any of the $k$ wallets of the anonymity set and from any proxy node these wallets are linked to. A wallet owner could use this fact to double-spend a token. For example, he could try to spend the same token from two different wallets and two different proxy nodes, breaking the double-spending protection property. To avoid this behavior, the $k$-Anon-AT TRANSFER operation is defined as an operation where a wallet owner $w$ asks a process $p$ to operate a transaction (labeled $T_i$). This transaction is a (ZK) proof that a given token exists and has not been spent. The transaction also transfers, by any means, the token to another wallet. The $k$ wallets (more precisely, the $n' \le k$ processes associated with the $k$ wallets) decide on a transaction order, i.e., they decide on the order of the pending transactions. Therefore, if a wallet owner tries to double-spend a coin, the network will only consider the first transaction. Finally, the transactions are forwarded to the rest of the network.

Let us extract knowledge from this definition. The TRANSFER validity property implies that the wallet owner can provide existence and non-double-spending proofs to the network. It implies that any other owner in the same anonymity set and with the same cryptographic material (randomness and associated element) can require the transfer of the same token.

We know the material required to produce a TRANSFER proof is stored in the wallet. Furthermore, we can assume that all the randomness used by a given wallet owner is produced by a randomness Oracle that derives a seed to obtain random numbers. Each seed is unique to each wallet. We assume the numbers output by an oracle seems random to an external observer, but two processes that share the same seed will obtain the same set of random numbers in the same order.

Finally, a transaction must be advertised to other processes and wallet owners via the TRANSFER operation. Therefore, proofs of transfer are public. We know these proofs are deterministically computed thanks to our deterministic random oracle model. Furthermore, only one sender and recipient are associated with each transfer operation. Therefore, the public proof cryptographically binds (without revealing them) the sender to the transaction. Hence, the public proof is a cryptographic commitment, which can be opened by the sender or any other actor who knows the same information as the sender.

In order to study the consensus number of this object, we consider that wallet owners can share their cryptographic material with the entire network, thereby giving up their anonymity. This would not make any sense in an anonymous system, but it represents a valuable tool to reason about the consensus number of the object. This sharing process can be implemented by an atomic register (and therefore has no impact on the consensus number, as we discuss later).

Processes can derive the sender's identity from the shared information using a local ``uncommit'' function. The ''uncommit'' function takes as input an oracle, a random seed, token elements, and an "on-ledger" proof of transfer of a token and outputs a wallet owner ID if the elements are valid. Otherwise, it outputs $\emptyset$.

% Any other actor can operate a transaction "on behalf" of $w$---this fact will allow us to prove the lower bound of the object type, but such an action could be called theft otherwise.

\subsection{Consensus number of the Anonymous Asset-Transfer object type}

\paragraph*{Lower bound}
\Cref{fig: Anon-AT lower} presents an algorithm that implements a $k$-consensus object, using only $k$-Anonymous Asset Transfer objects and SWMR registers. The $k$ in $k$-Anonymous Asset Transfer object refers here to the size of the biggest $\mu(w), \forall \ w\in \mathcal{W}$.

%\color{red}  Ajouter que l'on fait l'hypothèse que l'algo est mis en place par des owners, via des nodes, mais que on considère que les owners sont des nodes (avec l'abstraction: les nodes vérifient la validité d'un owner et signent pour re-créer un authenticated canal).

%Autre implication importante, ce n'est pas seulement l'anon-AT qui a besoin de synchro (entre les noeuds et le wallet owner), mais aussi n'importe quel système qui permet de dépenser depuis plusieurs noeuds différents. La justification est exactement la même. (prcque un w-o peut transmettre au même moment de plusieurs noeuds, pour empêcher le double spending, il faut que tout les noeuds, potentiellement envoyeur se synchonisent). -> L'algo résultant est le même, (je pense exactement, même la transmission de l'oracle et tt, même si ça c'est un peu moins sur).
%\color{black}

\begin{algorithm}
\begin{pchstack}[center,space=1em]
\resizebox{14cm}{!}{
\pseudocode[mode=text,codesize=\scriptsize]{
	\textbf{Shared variables}: \\
	\pcind AT $\leftarrow$ $k$-Anonymous-AT object, initialized with $k+1$ wallets, \\
	\pcind \pcind \pcind \pcind  each one of the $k$ first wallets possesses the elements \\
	\pcind \pcind \pcind \pcind necessary to transfer one shared token, the $k+1$-th \\
	\pcind \pcind \pcind \pcind wallet is the recipient of the transfers, it is not controlled\\
	\pcind \pcind \pcind \pcind by any process; \\
	\pcind RM-LEDGER $\leftarrow$ Atomic Snapshot object, initially $\{\emptyset\}^k$;\\
	\pcind V-LED $\leftarrow$ Atomic Snapshot object, initially $\{\emptyset\}^k$;\\
	\pcind O $\leftarrow$ A random oracle; \\
	\pcind TokenMat $\leftarrow$ secret associated with a unique token;\\
	\textbf{Local variables}: \\
	\pcind seed $\leftarrow$ random number;
}
\pseudocode[mode=text,codesize=\scriptsize]{
	\textbf{Operation} PROPOSE($v$) \textbf{is}: \\
	\pcln RM-LEDGER[p].update(seed, $p$);\\
	\pcln V-LED[p].update($v, p$);\\
	\pcln res $\leftarrow$ AT.transfer(TokenMat, O, seed, $k+1$);\\
	\pcln RML $\leftarrow$ RM-LEDGER.snapshot(); \\
	\pcln VL $\leftarrow$ V-LED.snapshot();\\
	\pcln \textbf{For} $i$ in $\{1, \cdots, k\}$ \textbf{do}:\\
	\pcln\pcind \textbf{If} uncommit(O, RML[$i$], TokenMat, res) $\ne \emptyset$ \textbf{then}:\\
	\pcln\pcind \pcind \textbf{Return} VL[$i$];\\
	\pcln \textbf{Return} False;
}
}
\end{pchstack}
%\Description{Implementation of a $k$-consensus object using $k$-Anon-AT objects}
\caption{Implementation of a $k$-consensus object using $k$-Anon-AT objects}
\label{fig: Anon-AT lower}
\end{algorithm}

\begin{theorem}
\Cref{fig: Anon-AT lower} wait-free implements $k$-consensus.
\end{theorem}

\begin{proof}
Let us fix an execution $E$ of the algorithm. Each PROPOSE operation only requires a finite number of AT-transfer operations and a finite number of accesses to Atomic Snapshot objects. Both objects are assumed to be atomic. The number of wallet owners is finite. Therefore, each process finishes the invocation of PROPOSE in a finite number of its own steps. Let $H$ be the history of $E$. We define $\bar{H}$ the completion of $H$, where:
\begin{itemize}
	\item Any invocation of PROPOSE in $H$ which does not reach line $3$ is completed with the line "return False";
	\item Any invocation of PROPOSE in $H$ that reaches line $3$ is completed with the lines $4$ to $8$.
\end{itemize}
We call operations that return false failed operations. The other operations are called successful operations. We verify that the completed history $\bar{H}$ follow the specification of a $k$-consensus object:
\begin{itemize}
	\item Any process that invokes a failed PROPOSE in $\bar{H}$ is a faulty process. The fact that this process cannot decide on a value does not impact the validity, the agreement, or the termination properties.
	\item Any invocation of a successful PROPOSE operation in $\bar{H}$ returns the value proposed by the unique process that successfully transferred the token associated with TokenMat. If a process reaches line $4$ at time $t$, then a unique TRANSFER succeeded at time $t'<t$. Hence, the uncommit operation returns $1$ at least and at most once, ensuring the Validity property. The agreement is ensured because no two processes can spend the same coin. Furthermore, the coin associated with TokenMat is transferred to a dummy account which cannot invoke TRANSFER operations. Therefore, the agreement property is verified.
\end{itemize}
All invocations of the PROPOSE operation in $\bar{H}$ follow the specification, and the algorithm presented in \Cref{fig: Anon-AT lower} is wait-free. In conclusion, the proposed implementation of a $k$-consensus object is linearizable.
\end{proof}

\begin{corollary}
	The consensus number of the Anon-AT object is at least $k$.
\end{corollary}

\paragraph*{Upper Bound}

We give an implementation of the Anon-AT object using only Atomic Snapshot objects, \blist objects, and \wlist objects. Each wallet owner can request a TRANSFER operation to $k$ different processes.
The proposed implementation uses disposable tokens that are either created at the initialization of the system or during the transfer of a token. When a token is destroyed, a new token can be created, and the new owner of the token is the only one to know the cryptographic material associated with this new token. In the following, we use the zero-knowledge version of the \blist and \wlist object types, where all set-(non-)membership proofs use a zero-knowledge setup. In addition, we use an \wlist object to ensure that a token exists (no ex-nihilo creation), and we use a \blist object to ensure that the token is not already spent (double-spending protection).

The underlying cryptographic objects used are out of the scope of this paper. However, we assume our implementation uses the ZeroCash \cite{zcash} cryptographic implementation, which is a sound anonymous asset transfer protocol. More precisely, we will use a high-level definition of their off-chain functions. It is important to point out that using the ZeroCash implementation, it is possible to transfer value from a pseudonymous asset transfer object to an anonymous one using a special transaction called ``Mint''. To simplify our construction, we assume that each wallet is created with an initial amount of tokens $v_0$ and that our object does not allow cross-chain transfers. We, therefore, have no ``Mint'' operation.

ZeroCash uses a TRANSFER operation called \textit{pour} that performs
a transfer operation destroying and creating the associated
cryptographic material. Here, we use a modified version of \textit{pour}
which does not perform the transfer or any non-local operation. It is
a black-box local function that creates the cryptographic material required
prove the destruction of the source token ($T_O$) and the creation of
the destination one ($T_R$). Our modified \textit{pour} function takes
as input the source token, the private key of the sender ($\sk_s$),
and the public key of the recipient ($\pk_r$): \textit{pour}($T_O,
\pk_r, \sk_s$)$\rightarrow tx$, $tx$ being the cryptographic material
that makes it possible to destroy $T_O$ and create $T_R$.

There might be multiple processes transferring tokens concurrently. Therefore, we define a deterministic local function ChooseLeader($\mathcal{A}, tx$), which takes as input any set $\mathcal{A}$ and a transaction $tx$, and outputs a single participant $p$ which invoked BL.PROVE($tx$).\footnote{In reality, the signature of chooseLeader would be more complicated as the function needs $T_O, pk_r, sk_s$ in addition to $tx$. These additional elements make it possible to uncommit $tx$, thereby matching the values of the PROVE operation with $tx.T_R$. Note that this does not pose an anonymity threat as this is a local function invoked by the owner of $sk_s$. We omit these details in the following to simplify the presentation.}

%ZeroCash is a fully anonymous Money transfer protocol. The underlying anonymity set is the set of all the network wallets. It can be made $k$-anonymous using multiple ZeroCash frameworks with only $k$ wallets. The recipient's anonymity set is defined by the registration \wlist of the newly created token.

\begin{algorithm}
\begin{pchstack}[center,space=1em]
\resizebox{14cm}{!}{
\pseudocode[mode=text,codesize=\scriptsize]{
	\textbf{Shared variables}: \\
	\pcind DL $\leftarrow$ $k$-\blist object, initially $(\emptyset, \emptyset)$; \\
	\pcind AL $\leftarrow$ \wlist object, initially $(\{(token_{(i,j)})_{i=1}^t\}^{k}_{j=1}, \emptyset)$\\
	%% \textbf{Local variables}: \\
	%% \pcind Local-Storage$_i$ $\leftarrow$ A local list of all tokens owned by the\\
	%% \pcind \pcind \pcind \pcind \pcind   wallet $i$, initially it contains the proofs of \\
	%% \pcind \pcind \pcind \pcind \pcind possession of $t$ tokens;\\
	%% \textbf{Local Function} LOCALBALANCE() is:\\
	%% \pcln  \textbf{Return} $|\mathrm{Local-Storage}|*v$;\\
	\textbf{Operation} TRANSFER($T_O, \pk_r, \sk_s$) \textbf{is}: \\
	\pcln  $tx \leftarrow$ Pour($T_O, \pk_r, \sk_s$)\\
	\pcln  \textbf{If} verify($tx$) and $tx \in$ AL and $tx \notin$ DL \textbf{then}:\\
	\pcln  \pcind AL.PROVE($tx$);\\
	\pcln  \pcind DL.PROVE($tx$);
}
\pseudocode[lnstart=4,mode=text,codesize=\scriptsize]{
	\pcln  \pcind DL.APPEND($tx$);\\
	\pcln  \pcind \textbf{Do}: \\
	\pcln  \pcind \pcind ret $\leftarrow$ DL.PROVE(tx);\\
	\pcln  \pcind\textbf{While} ret $\ne$ false;\\
	\pcln  \pcind \textbf{If} ChooseLeader(DL.READ(), $tx.T_R$)=$p$ \textbf{then}: \\
	\pcln  \pcind \pcind AL.append($tx.T_R$);\\
	\pcln  \pcind \pcind \textbf{Return} $tx.T_R$;\\
	\pcln  \textbf{Return} False;
}
}
\end{pchstack}

%\Description{Anon-AT object implementation using SWMR registers, \wlist objects, and \blist objects.}
\caption{Anon-AT object implementation using SWMR registers, \wlist objects, and \blist objects.}
\label{fig: Anon-AT upper}
\end{algorithm}

\begin{theorem}
	\Cref{fig: Anon-AT upper}  wait-free implements an Anon-AT object.
\end{theorem}

\begin{proof}
Let us fix an execution $E$ of the algorithm. $E$ only uses \blist objects and SWMR registers (\wlist objects have consensus number 1 and can be implemented using SWMR registers). Each TRANSFER operation only requires a finite number of \blist.PROVE and \blist.APPEND operations, which are assumed atomic. Each process finishes the invocation of TRANSFER in a finite number of its own steps. Let $H$ be the history of $E$. We define $\bar{H}$ the completion of $H$, where:
\begin{itemize}
	\item Any pending invocation of TRANSFER in $H$ that did not reached line $11$ is completed with the line "return False";
	\item Any pending invocation of PROPOSE in $H$ that reached line $11$ is completed with line $12$.
\end{itemize}
We call operations that return false failed operations. The other operations are called successful operations. In the following, we analyze if the completed history $\bar{H}$ follows the Anonymous Asset-Transfer object specification:

\begin{itemize}
	\item Any process that fails in an invocation of a TRANSFER operation in $\bar{H}$ is a faulty process. It cannot transfer money or create or double-spend a token. Therefore, it does not contradict the properties of the Anonymous Asset Transfer object. This faulty process can lose tokens (by destroying it and not transferring it), but because we assume a crash is definitive, all its tokens are lost anyway.
	\item A TRANSFER($T_O$, $T_R$) operation invoked by a process $p$ in $\bar{H}$ that succeeds proves to the network that the token $T_O$ exists and was not spent. After creating the cryptographic material for the new token and for destroying the old one, the operation verifies the correctness of the material and the validity of the old and new tokens (line 2). The prove operation in line 3 necessarily succeeds as $tx \in AL$. The prove in line 4 may instead fail if another process is performing a concurrent transfer operation for the same $T_O$. This potential conflict is resolved by the ChooseLeader function, which takes as input the list of successful prove operations on the \blist. The determinism of the ChooseLeader function and the agreement provided by the READ operation ensure that all processes add the same $tx.T_R$ at line 10 even if multiple processes issue concurrent conflicting TRANSFER operations for the same $T_O$.
          %% As said before, two transactions $tx$ can be added to the \wlist. It does not impact the validity of the object. Furthermore, it is impossible to append two different $tx$ values to the \wlist, thanks to the transitive verification ($tx$ must be proved in AL, then in DL, then appended in DL, and finally appended to AL).
The previous statement enforces double-spending prevention. The non-creation property is ensured by the PROVE operation conducted on the \wlist on line $3$. Finally, anonymity is enforced using the ZKP version of the \wlist and the \blist objects.
\end{itemize}
All invocations of the TRANSFER operation in $\bar{H}$ follow the specification of the Anonymous Asset Transfer object. In conclusion, the proposed algorithm is a wait-free implementation of the Anonymous Asset Transfer object.
\end{proof}

\begin{corollary}
The consensus number upper bound of a $k$-anon-AT object is $k$. Using this corollary and the previous one, we further deduct that $k$-anon-AT object has consensus number 1.
\end{corollary}
\section{E-vote system implementation using a \blist object}\label{sec:evote}
%\mg{Ajouter: Le $k$ de $k$-\blist est important ici. Si federated, alors $k$=1 (ex:poincheval), si voter= node, $k=1$ (ex: Electronic Voting with Fully Distributed Trust and Maximized Flexibility Regarding Ballot Design), $k$ est le nombre de votes concurrent possible}
% To motivate the formal definition of the \blist object type, this section presents the instantiation of an e-vote object using a \blist object.

In this section, we show that \blist objects can provide upper bounds on the consensus number of a complex objects. As an example, we study an e-vote system. An e-vote system must comply with the same properties as an ''in-person'' voting. An ''in-person'' voting system must ensure four security properties, two for the organizers and two for the voters. First, the organizers of the vote must ensure that each person who votes has the right to do so.
Second, each voter must vote only once. Third, a voter must verify that their vote is considered in the final count. Fourth, an optional property is voter anonymity. Depending on the type of vote, the voter may want to hide their identity.

We want to design a distributed e-vote system, where a voter can submit their vote to multiple different voting servers---while ensuring the unicity of their vote. We assume each server is a process of the distributed infrastructure. The voters act as clients, submitting vote requests to the servers. We assume the ''right to vote'' property is ensured using tokens. The Token is a one-time-used pseudonym that links a vote to a voter. Users obtains their Tokens from an issuer. All the voting servers trust this issuer. Neither the voters nor the issuer has access to the e-vote object---except if one of the actors assumes multiple roles. Using these specifications, we define the e-vote object type as follows:

\begin{definition}
The e-vote object type supports two operations: \emph{VOTE}(\textit{Token}, $v$) and \emph{VOTE-COUNT}(). \textit{Token} is the voter's identifier, and $v$ is the ballot. Moreover, these operations support three mandatory properties and one optional property:
\begin{enumerate}
	\item (Vote Validity) A \emph{VOTE}(\textit{Token}, $v$) invoked at time $t$ is valid if:
	\begin{itemize}
		\item \textit{Token} is a valid token issued by an issuer trusted by the voting servers; and
		\item No valid \emph{VOTE}(\textit{Token}, $v'$) operation was invoked at time $t'<t$, where $v'\ne v$.
	\end{itemize}
	\item (\emph{VOTE-COUNT} validity) A \emph{VOTE-COUNT}() operation returns the set of valid \emph{VOTE} operations invoked.
	\item (optional - Anonymity) A \textit{Token} does not link a vote to its voter identity, even if the voting servers and the issuers can collude.
\end{enumerate}
\end{definition}

In the following, we analyze an e-vote system based on signatures. The issuer issues a signature to the voter. The message of the signature is a nonce. The tuple (signature, nonce) is used as a token. When a voter issues a vote request, the server verifies the signature's validity and proceeds to vote if the signature is valid.

Adding anonymity to a signature-based e-vote system can be easily achieved using blind signatures \cite{chaum-blind-sig}.
A blind signature algorithm is a digital signature scheme where the issuer does not learn the value it signs. Because the signed value is usually a nonce, the issuer does not need to verify the value---a value chosen maliciously will not grant the voter more privileges than expected.

Formally, a Blind signature algorithm is defined by the tuple (Setup, Commit, Sign, Uncommit, Verify), where Setup creates the common values of the scheme (generators, shared randomness, etc...) and secret/public key pairs for all issuers. The public keys are shared with all the participants in the system. Commit is a commitment scheme that is hiding and binding; it outputs a commitment to a value---the nonce---randomly chosen by the user. The Sign algorithm takes as input a commitment to a nonce and the secret key of an issuer. It outputs a signature on the commitment. The Uncommit algorithm takes a signature on a commitment and the issuer's public key as input and outputs the same signature on the uncommitted message (the original message). Finally, the Verify algorithm outputs $1$ if the uncommitted signature is a valid signature by an issuer on the message $m$.

\Cref{fig: e-vote upper} provides a wait-free implementation of an e-vote system using any signature scheme, one $k$-\blist object, and one SWMR atomic snapshot object. Here, the value of $k$ corresponds to the number of voting servers, and $k$=$|\Pi_V|=|\Pi_M|$.
The idea of the algorithm is to use the APPEND operation to state that a token has already been used to cast a vote. In order to obtain a wait-free implementation, we use a helper value~\cite{help} stored in the Atomic Snapshot object AS-prevote. The vote operation is conducted as follows: the voting server $V$ communicates the vote it will cast in the AS-prevote object. Then, $V$ conducts a PROVE(\textit{Token}) operation to prove that the Token has not yet been used. Then, $V$ invokes an APPEND(\textit{Token}) operation and waits until the APPEND is effective---the do-while loop in line $8$ to $10$. Finally, $V$ uses the READ operation to verify that it is the only process that proposed a vote for this specific Token. If it is the case, the vote is added to the vote array---the AS-vote array. Otherwise, the vote is added to the vote array only if the other servers that voted using the same Token proposed the same ballot in line $4$.

Other implementations can be proposed in the case of two concurrent transactions requested by the same voter---to different servers---with different values. For example, it is possible to modify the algorithm presented in \Cref{fig: e-vote upper} using a deterministic function to choose one value among all the potential votes. This modification does not impact the properties of this implementation.

We now provide an informal proof of the linearizability of this implementation. The anti-flickering property of the $k$-\blist object ensures the termination of the while loop. Therefore, the implementation is wait-free. The same property ensures that the \textit{vote-values} variables are the same for all voters with the same Token, thus ensuring the unicity of the vote. The proof of authorization of the vote is ensured by the signature verification in line $1$. The anonymity property is also fulfilled if a blind signature scheme is used. Therefore, the use of anonymous \blist objects is not required. Hence, we can conclude that the consensus number of the $k$-\blist object type is an upper bound on the consensus number of an e-vote system.

\begin{algorithm}
\begin{pchstack}[center,space=1em]
\resizebox{14cm}{!}{
\pseudocode[mode=text,codesize=\scriptsize]{
	\textbf{Shared variables}: \\
	\pcind $k$-dlist $\leftarrow$ $k$-\blist object; \\
	\pcind AS-prevote $\leftarrow$ Atomic Snapshot object, initially $\{\emptyset\}^k$ \\
	\pcind AS-vote $\leftarrow$ Atomic Snapshot object, initially $\{\emptyset\}^k$ \\
	\textbf{Operation} VOTE((\textit{signature}, \textit{pk}, \textit{token}, $v$) \textbf{is}: \\
	\pcln  \textbf{If} Verify(\textit{signature}, \textit{token}, \textit{pk}) $ \ne 1$ \textbf{then}:\\
	\pcln  \pcind \textbf{Return} false;\\
	\pcln  AS-prevote.update((\textit{token}, $v$), $p$);\\
	\pcln  ret $\leftarrow$ $k$-dlist.PROVE(\textit{token})\\
	\pcln  \textbf{If} ret = false \textbf{then}:\\
	\pcln  \pcind \textbf{Return} false;\\
	\pcln  $k$-dlist.APPEND(\textit{token});\\
	\pcln  \textbf{Do}: \\
	\pcln  \pcind ret $\leftarrow$ $k$-dlist.PROVE(\textit{token});\\
	\pcln  \textbf{While} ret $\ne$ false;}
\pseudocode[lnstart=10, mode=text,codesize=\scriptsize]{
	\pcln  \textit{votes} $\leftarrow$ $k$-dlist.READ();\\
	\pcln  \textit{client-votes} $\leftarrow$ all values in \textit{votes} where token is \textit{token};\\
	\pcln  \textit{voters} $\leftarrow$ all processes in \textit{client-votes};\\
	\pcln  \textit{vote-values} $\leftarrow$ all values in AS-prevote\\
	\pcind \pcind \pcind \pcind `\pcind  where token is \textit{token} and processes that added the value are in \textit{voters}; \\
	\pcln  \textbf{If} \textit{vote-values} = $\{v\}^l, \forall\ l\ge1$ \textbf{then}:\\
	\pcln \pcind previous-votes $\leftarrow$ AS-vote.SNAPSHOT()[$p$];\\
	\pcln  \pcind AS-vote.UPDATE(previous-votes $\cup$ (token, $v$, voters), $p$);\\
	\pcln  \pcind \textbf{Return} true;\\
	\pcln  \textbf{Else return} false;\\
	\textbf{Operation} VOTE-COUNT() \textbf{is}: \\
	\pcln  votes $\leftarrow$ all values in AS-vote.Snapshot(). \\
	\pcind \pcind \pcind \pcind `\pcind Only one occurrence of each tuple (token, $v$, voters) is kept;\\
	\pcln  \textbf{Return} votes;
}
}
\end{pchstack}
\caption{Implementation of an e-vote object using one $k$-\blist object and Atomic Snapshots}
\label{fig: e-vote upper}
\end{algorithm}

It is also possible to build a wait-free implementation of a $k$-consensus object using one e-vote object. This implementation is presented in \Cref{fig: e-vote lower}. The idea of this algorithm is that each process will try to vote with the same Token. Because only one vote can be accepted, the e-vote object will only consider the first voter. Ultimately, every process will see the same value in the vote object. This value is the result of the consensus.

Each invocation is a sequence of a finite number of local operations and e-vote object accesses, which are assumed atomic. Therefore, each process terminates the PROPOSE operation in a finite number of its own steps. A vote can only be taken into account if it was proposed by some process, which  enforces the validity property. The agreement property comes from the unicity of the vote. Finally, the non-trivial property of the consensus object is ensured because the decided value is any value $v$ chosen by the winning process.
\begin{algorithm}
\begin{pchstack}[center,space=1em]
\pseudocode[mode=text,codesize=\scriptsize]{
	\\ \textbf{Shared variables}: \\
	\pcind vote-obj $\leftarrow$ e-vote object where the only authorized token is $0$; \\
	\pcind $\sigma \leftarrow$ signature on $0$ by a trusted issuer;\\
	\textbf{Operation} PROPOSE($v$) \textbf{is}: \\
	\pcln vote-obj.VOTE(($\sigma$, 0), $v$);\\
	\pcln $\{($winner-token, value$)\}$ $\leftarrow$ vote-obj.VOTE-COUNT();\\
	\pcln \textbf{Return} value;}
\end{pchstack}
%\Description{Implementation of a $k$-consensus object using one e-vote object}
\caption{Implementation of a $k$-consensus object using one e-vote object}
\label{fig: e-vote lower}
\end{algorithm}

The consensus number of a blind-signature-based e-vote system is bounded on one side by the consensus number of a $k$-consensus object and the other side by the consensus number of a $k$-\blist object. Hence, the blind-signature-based e-vote object type has consensus number $k$.

\end{document}